\documentclass[superscriptaddress,nofootinbib,notitlepage,preprintnumbers]{revtex4}
\pdfoutput=1

\usepackage{graphicx}				
\usepackage{amssymb}
\usepackage{amsmath}								
\usepackage{color}
\usepackage[utf8]{inputenc}
\usepackage{longtable}
\usepackage{placeins}
\usepackage{multirow}

\def\be{\begin{equation}}
\def\ee{\end{equation}}
\def\ba{\begin{array}}
\def\ea{\end{array}}

\DeclareMathAlphabet\Scr{U}{rsf}{m}{n} \makeatletter
\@addtoreset{equation}{section} \makeatother

\def\be{\begin{equation}}
\def\ee{\end{equation}}
\def\ba{\begin{array}}
\def\ea{\end{array}}

\newcommand{\bea}{\begin{eqnarray}}
\newcommand{\eea}{\end{eqnarray}}

\DeclareMathAlphabet \mathbfcal{OMS}{cmsy}{b}{n}

\begin{document}

\preprint{OUTP-17-13P}
\preprint{DESY 17-146}

\title{A Unified Model of Quarks and Leptons with a Universal Texture Zero}

\author{Ivo de Medeiros Varzielas}
\email{ivo.de@udo.edu}
\affiliation{CFTP, Departamento de F\'{i}sica, Instituto Superior T\'{e}cnico, Universidade de Lisboa, Avenida Rovisco Pais 1, 1049 Lisboa, Portugal}
\author{Graham G. Ross}
\email{g.ross1@physics.ox.ac.uk}
\affiliation{Rudolf Peierls Centre for Theoretical Physics,
University of Oxford,
1 Keble Road, Oxford,  OX1 3NP, U.K.}
\author{Jim Talbert} 
\email{james.talbert@desy.de}
\affiliation{Theory Group, Deutsches Elektronen-Synchrotron (DESY), D-22607 Hamburg, Germany}

\begin{abstract}We show that a universal texture zero in the (1,1) position of all fermionic mass matrices, including heavy right-handed Majorana neutrinos driving a type-I see-saw mechanism, can lead to a viable spectrum of mass, mixing and CP violation for both quarks and leptons, including (but not limited to) three important postdictions:  the Cabibbo angle, the charged lepton masses, and the leptonic `reactor' angle. We model this texture zero with a non-Abelian discrete family symmetry that can easily be embedded in a grand unified framework, and discuss the details of the phenomenology after electroweak and family symmetry breaking. We provide an explicit numerical fit to the available data and obtain excellent agreement with the 18 observables in the charged fermion and neutrino sectors with just 9 free parameters.  We further show that the vacua of our new scalar familon fields are readily aligned along desired directions in family space, and also demonstrate discrete gauge anomaly freedom at the relevant scale of our effective theory.
\end{abstract}

\maketitle

\pagebreak
\section{Introduction}
A convincing theory of fermion masses has proven elusive. Indeed, there is not even consensus as to whether the pattern of quark, charged lepton, and neutrino masses is determined by dynamics or is anarchical in nature.  This confusion is partially driven by measurements of highly incongruent masses and mixings between the quark and lepton sectors, and hence most attempts to explain this disparate spectrum choose a family symmetry that treats them differently.  There is little evidence for such extended structures given limited available data. 

In this paper we study whether, due to an underlying see-saw mechanism generating neutrino masses, a unified and family symmetric description of flavour is possible. Importantly, an enhanced symmetry of the system can lead to a universal texture zero (UTZ) in all fermion mass matrices (Dirac and Majorana) that gives quantitative postdictions for masses, mixings, and CP violation in good agreement with data, a significant advantage over anarchical schemes. 

The structure of the underlying family symmetry we choose to study is motivated by the observation that neutrino mixing is quite close to tribimaximal mixing, in which limit the neutrino mass eigenstates are given by  
\be
\begin{array}{l}
{\nu _a} = \left( {{\nu _\mu } + {\nu _\tau }} \right)/\sqrt 2 \\
{\nu _b} = \left( {{\nu _e} + {\nu _\mu } - {\nu _\tau }} \right)/\sqrt 3 \\
{\nu _c} = \left( {2{\nu _e} - {\nu _\mu } + {\nu _\tau }} \right)/\sqrt 6 
\end{array}
\label{eigenstates}
\ee

This pattern follows if there is a $Z_2\times Z_2$ discrete group closed, in the $\nu_{a,b,c}$ basis, by the generators\footnote{Acting with $S$ permutes $\nu_{a} \to -\nu_{a}$, $\nu_{b} \to -\nu_{b}$, $\nu_{c} \to \nu_{c}$ and similarly for $U$.} 
\[S = Diag(-1,-1,1),\;U = Diag(1,-1,-1)\] 
which form a subgroup of $S_4$ \cite{Lam:2007qc} or, more generally, of an $SU(3)$ family symmetry. In the pioneering works with $A_4$ \cite{Altarelli:2005yp, Ma:2005qf, Altarelli:2005yx, deMedeirosVarzielas:2005qg}, only one of the $Z_2$ is a subgroup of the family symmetry, the other $Z_2$ is generated accidentally due to the specific choice of representations of $A_4$. In any case, this $Z_2\times Z_2$ symmetry must be broken to obtain an acceptable value of the leptonic `reactor' angle $\theta_{13}^{l}$.  More significantly, the symmetry must be strongly broken in the quark and charged lepton sectors where the heaviest states are mainly aligned along the third generation, leaving an approximate $SU(2)$ symmetry.  To obtain a universal description of fermions it is therefore necessary that aspects of both the $Z_2\times Z_2$ and approximate $SU(2)$ symmetries be present.  Many attempts at integration have been made in the literature (see e.g. \cite{King:2017guk} for a recent review), including a host of model-independent scans of finite groups 
\cite{deAdelhartToorop:2011re, Lam:2012ga, Holthausen:2012wt, Holthausen:2013vba, King:2013vna, Lavoura:2014kwa, Talbert:2014bda, Joshipura:2014pqa, Joshipura:2014qaa, Yao:2015dwa, King:2016pgv, Varzielas:2016zuo,Yao:2016zev} which impose specific breaking patterns down to the desired `residual' symmetries in the quark and/or lepton sectors.  Unfortunately these scans only yield partially successful results for very large groups, and in any event do not attempt to explain the dynamics of the purported symmetry breaking. 

To achieve the desired patterns of mass and mixing we instead consider the group $\Delta(27)$ \cite{deMedeirosVarzielas:2006fc, Ma:2006ip, Luhn:2007uq, Varzielas:2015aua}, which can be understood as the $(Z_3\times Z_3) \rtimes Z_3$ semi-direct product symmetry (see \cite{Ishimori:2010au} for a detailed discussion of the group properties of $\Delta(27)$). We note that the $Z_2\times Z_2$ neutrino symmetry above is not a subgroup of $\Delta(27)$; it appears indirectly due to the specific vacuum alignments which arise naturally from $\Delta(27)$.  Family symmetry eigenstates transform as  $\psi_j\rightarrow e^{i\alpha_j}\psi_j$ ($j=1,2,3$) under the first $Z_3$ with $\alpha_j = 2\pi j/3$ and as cyclic permutations under the last $Z_3$.  As we discuss in the Appendix, spontaneous symmetry breaking of either $Z_3$ can readily occur through triplet familon fields acquiring vacuum expectation values (vevs). For the case the vev is in the $\left\langle {{\theta _3}} \right\rangle  \propto (0,0,1)$ direction the symmetry is broken to the $Z_3$ phase symmetry.  Coupling of the $\theta_3$ familon to fermions can lead to fermion masses for the third generation. For the case the vev is in the $\left\langle {{\theta _{123}}} \right\rangle  \propto (1,1, - 1)$ direction, the symmetry breaks to the $Z_3$ permutation symmetry (up to a rephasing, see Appendix \ref{VEV_align}) and a mass can be generated for a combination of fermion generations, such as $\nu_b$, along this direction. If there are several familons both types of vev can arise and indeed a further familon can acquire vevs along the $\left\langle {{\theta _{23}}} \right\rangle  \propto (0,1,1)$ direction, allowing for mass generation for fermions such as the $\nu_a$ state.

Given this pattern of symmetry breaking (and including an additional cyclic shaping symmetry) it is possible to have a ubiquitous structure for the Dirac matrices describing up and down quarks, charged leptons, and neutrinos. This gives a UTZ in the $(1,1)$ direction that results in excellent postdictions for quark and charged lepton masses. The light neutrino mass and mixing structure is quite different because the right-handed neutrinos can also have large Majorana masses and, for the case this is dominated by the third generation mass, sequential dominance takes place \cite{Smirnov:1993af,King:1998jw, King:1999cm, King:1999mb, King:2002nf}  and the see-saw mechanism suppresses the large third generation Dirac mass matrix contribution, allowing for the light neutrino mass eigenstates to be approximately given by eq(\ref{eigenstates}). However, unlike previous models of this type (see e.g. \cite{Bjorkeroth:2015ora, Bjorkeroth:2015uou, Bjorkeroth:2017ybg} and references therein), the $(1,1)$ texture zero structure remains after the see-saw and leads to a specific departure from pure tribimaximal mixing in the neutrino sector, and thus gives a non-zero $\theta_{13}^{l}$.

The organisation of the paper is as follows: In Section \ref{Dirac mass matrix} we discuss the Dirac matrix structure needed to get acceptable masses and mixings for the charged fermions, concentrating on the maximally symmetric form. We show how, with an additional shaping symmetry, the structure can be generated by coupling the fermions to the $\theta_3,\;  \theta_{23}$ and $\theta_{123}$ familons. In Section \ref{neutrinos} we use the underlying symmetries to constrain both the Dirac and Majorana mass matrices for the neutrinos, show that the $(1,1)$ texture zero persists for the light neutrino mass matrix after a type-I see-saw, and discuss the generic form of the resulting relations between masses and mixings. In Section \ref{anomaly} we explore the consistency of our model when continued to the UV by discussing the relevant discrete gauge anomalies present, and ultimately show that our model as currently formatted is safe. Finally in Section \ref{Pheno} we show that, with a reasonable choice of the  parameters of the model, a quantitatively  acceptable structure for the  masses and mixing of quarks and leptons results. The details of vacuum alignment are presented in the Appendix.

\section{Charged fermion mass structure}\label{Dirac mass matrix}

An important issue in the determination of fermion mass predictions is the scale at which the prediction applies. If this is large, at the GUT or Planck scale, there will be significant radiative corrections which depend on the structure of the theory up to this scale. In this paper we assume the scale is indeed large and further that supersymmetry (SUSY) \footnote{Our flavour model does not necessarily rely on the specifics of the MSSM, and indeed the additional familons we employ are not part of its spectrum, but of course the physical parameters we study (and ultimately fit) must be radiatively corrected to the UV in a model-dependent way.  To do this we assume the MSSM does play a role in our vacuum alignment, cf.  Appendix \ref{VEV_align}.} prevents radiative corrections from driving an unacceptably large electroweak breaking scale (the hierarchy problem) and allows for precise gauge coupling unification.  In this case the radiative corrections due to gauge interactions are well understood for the quarks. Due to the fact that the QCD coupling is much larger than the electroweak couplings, at low scales the quark masses are enhanced by about a factor of 3 relative to the lepton masses.  However this enhancement is reduced by renormalisation group flow generated by Yukawa couplings and this introduces considerable uncertainty due to the fact that in SUSY the top and bottom Yukawa couplings depend sensitively on $\tan\beta$, the ratio of the vevs of the two Higgs doublets in the MSSM \cite{Olechowski:1990bh, Ross:2007az, Chiu:2016qra}. In addition there may be large SUSY threshold corrections \cite{Ross:2007az}.

Taking these corrections into account, quark and charged lepton masses and mixings are consistent with  a symmetric\footnote{We are interested in the maximum symmetry consistent with all fermion masses and mixing - hence the choice of symmetric mass matrices.  An underlying $SO(10)$ symmetry may be the origin of this structure.} mass matrix structure of the form
\be
\renewcommand{\arraystretch}{1.3}
M_a^D\approx m_3 \left( {\begin{array}{*{20}{c}}
0&{\varepsilon _a^3}&{  \varepsilon _a^3}\\
{\varepsilon _a^3}&{{r_a}\varepsilon _a^2}&{  {r_a}\varepsilon _a^2}\\
{  \varepsilon _a^3}&{  {r_a}\varepsilon _a^2}&1
\end{array}} \right),\quad {r_{u,d}}=1/3,\quad {r_e} =   -1
\label{mm1}
\ee
This describes the observed masses and mixings provided the parameters $\epsilon_a, \; a=u,d,e$ differ between the up quark and down quark/charged lepton sectors with $\epsilon_u\approx 0.05,\;\epsilon_{d,e}\approx 0.15$. 
This symmetric structure has a (1,1) texture zero and, in the quark sector, implements the Gatto-Sartori-Tonin relation \cite{Gatto:1968ss} for the Cabibbo angle given by 
\be \sin{\theta _c} = \left\vert \sqrt {\frac{{{m_d}}}{{{m_s}}}}  - {e^{i\delta }}\sqrt {\frac{{{m_u}}}{{{m_c}}}} \right\vert
\label{GST}
\ee
for some phase $\delta$. With $\delta\approx \pi/2$ this is in excellent agreement with the observed masses and mixing angle. The factors $r_i$  implement the Georgi-Jarlskog mechanism \cite{Georgi:1979df} giving  ${m_b} = {m_\tau },\;{m_\mu } = 3{m_s},\;{m_e} = {\textstyle{1 \over 3}}{m_d}$  at the unification scale, which is also in good agreement with the measured values after including radiative corrections \cite{Ross:2007az}.

\subsection{Familon description}
This structure can be obtained by coupling the fermions to familons $\theta_i$, provided the discrete family symmetry is supplemented by  an underlying shaping symmetry.  In writing the effective Lagrangian preserving the underlying discrete $\Delta(27)$ symmetry we assume that only triplet representations are present and that the higher dimensional operators that arise are just those consistent with the exchange of triplets, ensuring that, at the non-renormalizable level, there are no contractions involving the non-trivial singlets of $\Delta(27)$.\footnote{This structure is found in orbifold string compactifications \cite{Nilles:2012cy}.} The difference between the down quark and charged lepton matrices can be derived from an underlying GUT structure. As an example of this consider the effective Lagrangian of the form
\be 
\mathcal{L}^{eff}_{a,mass}=\psi_i\left ({1\over M_{3,a}^2}\theta_3^i  \theta_3^j  +{1\over M_{23,a}^3}\theta_{23}^i\theta_{23}^j\Sigma+{1\over M_{123,a}^3}(\theta_{123}^i\theta_{23}^j+\theta_{23}^i\theta_{123}^j)  S    \right)\psi_j^{c}H_5 \label{mass}
\ee
where $a=u,\;d,\;e$ and
\be
\left\langle {{\theta _3}} \right\rangle  = {{\rm{v}}_3}(0,0,1),\;\left\langle {{\theta _{23}}} \right\rangle  = {{\rm{v}}_{23}}(0,1,1)/\sqrt 2 ,\;\left\langle {{\theta _{123}}} \right\rangle  = {{\rm{v}}_{123}}(1,1, - 1)/\sqrt 3 
\ee
The restricted form of eq(\ref{mass}) is determined by a simple $Z_N$ shaping symmetry under which the fields with non-zero $Z_N$ are shown in Table \ref{tab:Zcharges}, along with the full symmetry assignments of our model. The field $S$ is $Z_N$ charged and indirectly affects the Majorana terms such that the UTZ is preserved (see Section \ref{neutrinos}). The field $\Sigma$ is associated with the breaking of the underlying GUT with a vev $\propto B-L+\kappa T_3^R$.  It implements the Georgi-Jarlskog relation \cite{Georgi:1979df} with ${r_e / r_d}=-3$ for $\kappa=0$. For the case $\kappa=2$, plus domination by the RH messengers, it gives ${r_e / r_d}=3$. Since the sign is irrelevant both cases are viable.  Here we concentrate on the case $\kappa=0$ which gives $r_\nu=-1$ and $r_u/r_d=1$. 
We note that, although we do not go into the details of the GUT breaking, we checked it can proceed as normal from an underlying $SO(10)$ down to the SM gauge group. The reasons for this are that the $H_5$ field that breaks $SO(10)$ to the Pati-Salam group is neutral under the $Z_N$ and that, although $\Sigma$ has a non-trivial $Z_N$ charge, it can obtain a VEV from non-holomorphic terms in the potential that are traces of the ($Z_N$ invariant) combination $\Sigma \Sigma^\dagger$, which arise due to SUSY breaking, similarly to the terms responsible for the alignment of the familon VEVs discussed in more detail in Appendix \ref{VEV_align}.
Finally, the $M_{i,a}$ are the heavy masses of the mediators that have been integrated out when forming the effective Lagrangian. There is a subtlety in that at least the top Yukawa coupling should not be suppressed and to do this one must take $\theta_3/M_3$ large, a known issue in this type of model \cite{deMedeirosVarzielas:2005ax}. This is the case if  $\theta_3$ is the dominant contribution to the messenger mass, and we assume here that this applies to the $u,d$ and $e$ sectors. An alternative that solves this issue is through the use of Higgs mediators as described in \cite{Varzielas:2012ss}, although this is beyond the scope of the present paper as it requires an entirely different set of superfields.

\renewcommand{\arraystretch}{1.5}
\begin{table}[t]
\centering
\begin{tabular}{|c|c|c|c|c|c|c|c|c|c|c|}
\hline
$\text{Fields}$ & $\psi_{q,e,\nu}$  & $\psi^{c}_{q,e,\nu}$ & $H_{5}$ & $\Sigma$ & $S$ & $\theta_{3}$ &$\theta_{23}$ &$\theta_{123}$ & $\theta$ & $\theta_{X}$ \\
\hline
\hline
$\Delta(27)$ & 3 & 3 & $1_{00}$ & $1_{00}$ & $1_{00}$ & $\bar{3}$ & $\bar{3}$ & $\bar{3}$ & $\bar{3}$ & $3$  \\
\hline
$Z_{N}$ & 0 & 0 & 0 & 2 & -1 & 0 & -1 & 2 & 0 & $x$  \\
\hline
\end{tabular}
\caption{Fields and their family symmetry assignments.  The field $\theta_{X}$ only plays a role in the vacuum alignment.  Hence the only requirement of its $Z_{N}$ charge is that it be assigned so that the field does not contribute significantly to the fermionic mass matrices -- we have therefore left it generic.}
\label{tab:Zcharges}
\end{table}

\subsection{Mass matrix parameters and messenger masses}

The parameters of eq({\ref{mm1}) in the (2,3) block are given by
\be
\epsilon_a^2= {\langle\theta_{23}\rangle^2\langle\Sigma\rangle\over M_{23,a}^3}.{M_{3,a}^2\over \langle\theta_3\rangle^2}
\ee
Referring to the $Z_N$ charges of the fields as $Q$, if the $Q=0/Q=-1$ mediator mass ratio $M_3,a\over M_{23,a}$ is smaller in the up sector than in the down sector, one will have $\epsilon_u <\epsilon_d$. 
Of course equality of the down quark and charged lepton matrix elements in the (1,2), (2,1), and (3,3) positions requires that the expansion parameters be the same in the two sectors. This is consistent with an underlying spontaneously broken $SU(2)_R$ symmetry because the down quarks and leptons are both $T_{R,3}=-1/2$ states and, in SUSY, both acquire their mass from the same Higgs doublet, $H_d$.

Here we consider the case that the messengers carry quark and lepton quantum numbers. For the messengers carrying left-handed quantum number, $SU(2)_L$ requires the up and down messenger masses should be equal. Thus the only way the expansion parameters can be different in the up and down sectors is if the right-handed messengers dominate. In this case, if the underlying symmetry breaking pattern is 
\be
SO(10)\rightarrow  SU(4)\times SU(2)_L\times SU(2)_R\rightarrow SU(3)\times  SU(2)_L\times U(1)
\ee
the down quarks and charged leptons will have the same expansion parameter after $SU(2)_R$ breaking.

Up to signs  and $O(1)$ coefficients allowed by the $Z_N$ symmetry, the (1,j), (j,1) entries of eq(\ref{mm1}) are given by 
\be
\epsilon_{a}^{3} = {\langle \theta_{23} \rangle \langle \theta_{123} \rangle \langle S \rangle \over M_{123,a}^3}.{M_{3,a}^2\over \langle \theta_3 \rangle^2} ,
\ee
to be consistent with the form of eq(\ref{mm1}). Since they involve both the $Q=1$ and $Q=-1$ mediator masses there is sufficient freedom for this to be the case.  
\subsection{Higher order operators}
\label{sec:HigherOrder}
We may also be sensitive to terms of higher mass dimension in the operator product expansion of the effective theory.  The higher order operators allowed by the symmetries in Table \ref{tab:Zcharges} at the next order relevant to contributions in eq(\ref{mm1}) are of dimension eight, and hence are suppressed by four powers of the relevant messenger masses: 
\be
 \label{eq:HOLag}
\mathcal{L}^{HO}_{a,mass} = \psi_i\left ({1\over M_{23,a}^4}(\theta_{23}^i\theta_{3}^j + \theta_{3}^i\theta_{23}^j)\Sigma S+{1\over M_{123,a}^4}(\theta_{123}^i\theta_{3}^j+\theta_{3}^i\theta_{123}^j)  S^2    \right)\psi_j^{c}H_5
\ee
However, the relative magnitude of their contributions are of different orders in the mass matrix.  Assuming approximately universal messenger masses and given that the lowest order contributions involving the vev of $\Sigma$ are parametrically larger than those involving $S$, one finds that
\be
\frac{\langle \theta_{23} \rangle \langle \theta_{23} \rangle \langle \Sigma \rangle}{M_{23}^{3}} \sim \mathcal{O}(\epsilon^{2}), \,\,\,\,\,\,\,\frac{\langle \theta_{23} \rangle \langle \theta_{123} \rangle \langle S \rangle}{M_{123}^{3}} \sim \mathcal{O}(\epsilon^{3}) \,\,\,\Longrightarrow \,\,\, \frac{\langle \theta_{23} \rangle \langle \Sigma \rangle}{\langle \theta_{123} \rangle \langle S \rangle} \sim \mathcal{O}\left( \frac{1}{\epsilon} \right)
\ee  
where $\epsilon$ is the small parameter of eq(\ref{mm1}).  The contributions $\propto \Sigma S$ in eq(\ref{eq:HOLag}) are therefore also parametrically larger than those $\propto S^{2}$:
\be
\label{eq:HOcomp}
\frac{\langle \theta_{3} \rangle \langle \theta_{23} \rangle \langle \Sigma \rangle \langle S \rangle}{M^{4}} \sim \frac{1}{\epsilon} \frac{\langle \theta_{3} \rangle \langle \theta_{123} \rangle \langle S \rangle^{2}}{M^{4}} 
\ee
Hence we neglect the contributions to the mass matrix generated by the $S^{2}$ terms in the numerical fits performed in Section \ref{Pheno}.
Beyond the two terms discussed in this Section, the remaining higher order operators allowed by the symmetries have at least three additional insertions of $\Delta(27)$ triplets, and their contributions are negligible.

\section{Neutrino mass structure}\label{neutrinos}

The neutrino sector is not as well understood as the charged fermions, as only two mass-squared differences $\Delta m_{ij}^{2}$ and the three leptonic mixings angles $\theta_{ij}^{l}$ are constrained to a reasonable accuracy.  A recent global fit to available neutrino data from the NuFit Collaboration \cite{Nufit,Esteban:2016qun} finds $\Delta m^{2}_{21} \simeq 7.5 \times 10^{-5} \text{eV}^2$, $\Delta m^{2}_{31} \simeq 2.524 \times 10^{-3} \text{eV}^2$ (central values, normal mass ordering), and a leptonic PMNS mixing matrix within the $3\sigma$ confidence level of    
\begin{equation}
\label{eq:Nufit}
|V_{PMNS}|^{3\sigma} \in
\left(
\begin{array}{ccc}
(0.800 - 0.844) & (0.515 - 0.581)  & (0.139 - 0.155) \\
(0.229 - 0.516)   & (0.438 - 0.699) & (0.614 - 0.790)\\
(0.249 - 0.528)  & (0.462 - 0.715) & (0.595 - 0.776)
\end{array}
\right)
\end{equation}
The leptonic CP violating phase is not constrained at the $3\sigma$ confidence level.  Unlike quark mixing, leptonic mixing is clearly large, non-hierarchical, and still consistent with tribimaximal mixing up to obvious corrections in the (1,3) element.  However, neutrinos' fundamental nature as either Dirac or Majorana fermions, mass generation mechanism, absolute mass values, and associated CP violating phase(s) are currently unknown.  Furthermore, as with the charged fermions, we must be concerned about radiative corrections to neutrino mass and mixing parameters. The case where neutrino masses are generated with a type-I see-saw mechanism and radiatively corrected with an MSSM spectrum is well studied \cite{Casas:1999tg,Gupta:2014lwa,Casas:1999ac,Chankowski:1999xc,Antusch:2003kp}.  The authors of \cite{Antusch:2003kp} conclude that, while a degenerate (or nearly degenerate) mass spectrum, large $\tan{\beta}$, and/or special configurations of Dirac and Majorana CP violating phases can conspire and contribute to substantive running for the mixing parameters, the general expectation is that $\Delta \theta^{\nu}_{ij} \equiv \theta^{\nu}_{ij}(\Lambda_{GUT}) - \theta^{\nu}_{ij}(\Lambda_{MZ}) \sim \mathcal{O}(10^{-1} -10^{-3})$, even for rather large values of $\tan \beta$. Given that we predict a hierarchical mass spectrum with the lightest neutrino mass many orders of magnitude smaller than the rest, we take the current 3$\sigma$ bounds from NuFit to be valid in the UV as well.

Neutrino masses are more sensitive to radiative effects and can change by tens of percent over many decades of evolution to the UV.  In fact, in certain scenarios a normal spectrum in the UV can look like an inverted spectrum in the IR \cite{Antusch:2003kp}! Our solutions in the charged fermion sector tend to favor larger values of $\tan{\beta}$, and in this scenario the heaviest mass eigenstate will split from the lighter ones during its RGE.  This means that our principal mass prediction, the ratio of the solar and atmospheric mass splitting, will diminish in the UV.  Using the most recent values from NuFit one finds (in the IR) that
\begin{equation}
\frac{\Delta m^{2}_{sol}}{\Delta m^{2}_{atm}} \in \lbrace .0266, .0336\rbrace
\end{equation}
although we estimate that $\frac{\Delta m^{2}_{sol}}{\Delta m^{2}_{atm}} \gtrsim .021$ at the GUT scale, given the above discussion.

\subsection{Familon description}
We again find that this generic structure can be understood by coupling neutrino family triplets to familons although, due to the see-saw mechanism, the neutrino mass matrix will obviously have a different structure than the charged fermions. In the context of an underlying $SO(10)$ the neutrinos must have the  {\it same} form of the Dirac Lagrangian, eq(\ref{mm1}). Taking the case $\kappa=0$ gives $r_\nu=-1$. 

On the other hand, the  Majorana mass matrix requires lepton number violation. In the context of the familon structure introduced above it is an obvious choice to assume that the lepton number violation occurs through the vev of a further familon triplet field $\theta$ carrying lepton number $-1$. Then the Lagrangian terms responsible for the Majorana mass, consistent with the underlying $\Delta(27)$ symmetry, are given by\be
\mathcal{L}^{\nu}_{Majorana \;mass}=\psi_i^c \left ({1\over M}\theta^i  \theta^j  +{1\over M^4} [c_{1}\theta_{23}^i\theta_{23}^j(\theta^a\theta^a\theta_{123}^a)+c_{2}(\theta_{23}^i\theta_{123}^j+\theta_{123}^i\theta_{23}^j)(\theta^a\theta^a\theta_{23}^a)]\right )\psi^c_j
\label{MM}
\ee

Due to the different mediators (and couplings) we have allowed for different coefficients $c_{1}$, $c_{2}$ of the two components of the second term. In this form, we note the absence of terms with two $\theta_{123}$ familons, which would destroy the UTZ (the field $S$ which appears in the Dirac terms only is indirectly responsible for this absence).
The higher order operators allowed by the symmetries have at least three additional insertions of fields, and their contributions are again negligible. The lowest order operator with two $\theta_{123}$ familons in particular, appears with one additional $\theta_{23}$ familon and $S^3$.

 \subsection{Qualitative analysis of neutrino masses and mixing}
 \label{sec:nuqual}
 
The high inverse power of the mediator mass associated with the second term of eq(\ref{MM}) allows the hierarchical structure in the Majorana mass matrix to readily be much greater than that in the Dirac matrix. In this case the contribution to the LH neutrino masses via the see-saw with $\nu^c_3$ exchange is negligible and thus the mass matrix structure giving mass via the see-saw to the 2 heaviest neutrinos is effectively two dimensional. The Majorana mass matrix is defined in the $(\nu_1,\nu_2)$ basis and the Dirac mass matrix is in the  $(\nu_{b},\nu_{a})(\nu_1,\nu_2)$ basis where $\nu_{a,b}$ are given in eq(\ref{eigenstates}).
In this basis (and taking $\kappa = 0$) the application of the type-I see-saw generates a simple matrix of two complex parameters:
\be
M_{Majorana}\propto\left( {\begin{array}{*{20}{c}}
0&{c_{2}}\\
{c_{2}}&{{c_{1}+2\,c_{2}}}
\end{array}} \right),
\,\,\,\,\,\,
M_{Dirac}\propto\left( {\begin{array}{*{20}{c}}
0&{\sqrt{3/2}}\\
{1}&{{1+s}}
\end{array}} \right)\,\,\, \underbrace{\Longrightarrow}_{\text{see-saw}} \,\,\,
M_{\nu}\propto\left( {\begin{array}{*{20}{c}}
0&{-\sqrt{3/2}\,c_{2}}\\
{-\sqrt{3/2}\,c_{2}}& c_{1}^{\prime}
\end{array}}\right)
\label{eq:2Dseesaw}
\ee
where $c_{1}^{\prime} \equiv c_{1} - 2 \, c_2 \, s$ with $c_{1} \gg c_{2}$ and $s \propto \langle \Sigma \rangle \langle \theta_{23} \rangle/ (\langle S \rangle \langle \theta_{123} \rangle)$.  From this one easily finds that the ratio of neutrino masses is given by
\be
{m_2\over m_1}\approx {3\over 2} {c_{2}^2\over c_{1}^{\prime,2}},\,\,\,\,\,{c_{2} \over c_{1}^{\prime}}\equiv |{c_{2} \over c_{1}^{\prime}} |\,e^{i\eta},
\ee
defining the phase $\eta$, and that the heaviest neutrino mass eigenstate is
\be 
 \nu_1\propto \nu_{a}-e^{i\eta}\sqrt{{m_2\over m_1}}\,\nu_{b}
 \label{eq:heavyestate}
  \ee
Thus the (1,1) texture zero gives rise to the following mixing sum rules:
 \begin{align}
 \label{eq:13sumrule}
 \sin \theta^{\nu}_{13} &\approx \sqrt{{m_2\over 3m_1}} \\
  \label{eq:23sumrule}
 \sin\theta^{\nu}_{23} &\approx \vert {1\over \sqrt{2}}-e^{i\eta}\sin\theta^{\nu}_{13} \vert \\
 \label{eq:12sumrule}
 \sin\theta^{\nu}_{12} &\approx \frac{1}{\sqrt{3}}
 \end{align}
where the $\nu$ label indicates that only the contribution from the neutrino mixing matrix has been included.  Apart from the solar angle $\theta_{12}^{\nu}$, it is clear that the mixing deviates from the tribimaximal form, but now with too large a value for the reactor angle after inputing explicit experimental values for $m_{1,2}$ in eq(\ref{eq:13sumrule}).  We will show in Section \ref{Pheno} that an excellent value for $\theta_{13}^{l}$ is obtained after including the contributions predicted from the charged lepton sector, which also affect the solar and atmospheric mixing angles.  While we focus on an exact numerical approach in this paper, a detailed analytic discussion of these effects, including the relationship between $\eta$ and the standard Dirac CP violating phase $\delta_{CP}$, may be found in \cite{Hall:2013yha}.  
  
\section{Discrete Gauge Anomalies}  
\label{anomaly}
\begin{table}[t]
\centering
\begin{tabular}{|c|c|c|}
\hline
$\Delta(3N^{2})$ & $\bf{1}_{k,l}$ & $\bf{3}_{[k][l]}$ \\
\hline
$\det(h_{2})$ & $\omega^{k}$ & 1 \\
\hline 
$\det(h_{1})$ &  $\omega^{l}$ & 1 \\
\hline
$\det(h_{1}^{\prime})$ &  $\omega^{l}$ & 1 \\
\hline
\end{tabular}
\caption{Determinants over the generators of $\Delta(3N^{2})$ where $N/3 \in \mathbb{Z}$, for all irreducible representations of the group.  $\omega$ is the cubic root of unity, $\omega^{3} = 1$, while $h_{1}$, $h_{1}^{\prime}$ and $h_{2}$ simply denote the generators of the group. Finally, the $k,l$ indices simply indicate different irreducible representations -- see \cite{Ishimori:2010au} for a detailed discussion of the group properties of $\Delta(27)$.}
\label{tab:groupreps}
\end{table}
A long-standing argument of Krauss and Wilzcek \cite{Krauss:1988zc} holds that apparent global discrete symmetries  (Abelian $Z$ or non-Abelian $D$), e.g. R-Parity in standard SUSY models or our family symmetries, \emph{must} be local/gauged in order to avoid complications with quantum gravity (wormhole) effects.  Such discrete gauge symmetries should be anomaly free and the resultant constraints for the case of Abelian discrete symmetries were determined in \cite{Ibanez:1991hv,Ibanez:1991wt,Banks:1991xj}.  The analogous computation for non-Abelian discrete symmetries has since been formalized \cite{Araki:2007zza,Araki:2008ek,Ishimori:2010au} with a path-integral approach,\footnote{We use the notation of \cite{Araki:2008ek} in the equations that follow.} concluding that the only relevant anomalies in the IR assuming a fully massless spectrum are mixed non-Abelian gauge ($G$) and mixed gravitational ($g$) anomalies:
\begin{equation}
D-G-G, \,\,\,\,\,\,\,\,\,\, D-g-g,  \,\,\,\,\,\,\,\,\,\, Z-G-G, \,\,\,\,\,\,\,\,\,\, Z-g-g
\end{equation}
There are no IR anomaly constraints of the form  $\left[Z \right]^{2} U(1)_{Y}$ and $\left[ U(1)_{Y}\right]^{2} Z$ because the corresponding discrete charge $\alpha$ of any group element transformation is always defined modulo $N$, the order of the group element of the transformation, and as the hypercharges of the $U(1)$ symmetry groups can always be rescaled, one can do so such that this modulo constraint is satisfied. 

Furthermore, cubic discrete anomalies and mixed discrete anomalies of the form $Z-D-D$ or $D-Z-Z$ can be avoided by arguing charge fractionalization in the massive particle spectrum \cite{Ibanez:1991hv,Ibanez:1991wt,Araki:2008ek,Banks:1991xj,Csaki:1997aw}.\footnote{Failure to satisfy the cubic constraints can give valuable information about the ultimate order required of the $Z$ and/or $D$ groups.}  

The authors of \cite{Araki:2007zza,Araki:2008ek,Ishimori:2010au} conclude that the only difference between calculating the anomaly coefficient for an Abelian $Z_{N}$ or non-Abelian $D$ discrete symmetry is that, in the latter case, one must calculate the Abelian coefficients $Z_{N^{i}}$ for each generator $h_{i}$ of $D$.  We call the matrix representations of these elements $U$, and they live in some irreducible representation of $D$ labeled by $\bold{d}^{(f)}$:  
\begin{equation}
\label{NAelement}
U(\bold{d}) = e^{i \alpha(\bold{d})} = e^{i 2 \pi \,\tau(\bold{d})/N} 
\end{equation}
The condition for a discrete symmetry transformation to be anomaly-free is not uniquely determined, but is instead only determined modulo $N^{i}$.  Via a standard derivation, one can simultaneously read off the constraint on the anomaly coefficient for $Z-G-G$ or $D-G-G$:
\begin{equation}
\label{eq:zanom}
Z/D-G-G: \,\,\,\,\,\,\,\,\,\, \underset{\bold{r}^{(f)},\bold{d}^{(f)}}{\sum} tr\left[\tau(\bold{d}^{(f)})\right] \cdot l(\bold{r}^{(f)}) \overset{!}{=} 0 \,\, \text{mod}\,\, \frac{N}{2}
\end{equation}
The notation is such that the summation is only over chiral fermions living in representations that are non-trivial with respect to both $G$ and $D$.  $l(\bold{r}^{(f)})$ is the Dynkin index for a fermion living in a representation $\bold{r}^{(f)}$ of the gauge group.  It is normalized such that $l(M) = 1/2, 1$ for $\text{SU}(M)$ and $\text{SO}(M)$ respectively. Of course, Abelian discrete symmetries only have singlet irreducible representations. Here it is clear that $tr\left[\tau(\bold{d}^{(f)})\right]$ is a charge (called $\delta^{(f)}$ in \cite{Araki:2008ek}), and from eq(\ref{NAelement}) one notes that it can be written in terms of a (multi-valued) logarithm:
\begin{equation}
\label{NAcharge}
tr\left[\tau(\bold{d}^{(f)})\right] = N \frac{\text{ln}\, \text{det}\, U(\bold{d}^{(f)}) }{2 \pi i}
\end{equation}
For the Abelian case, $tr\left[\tau(\bold{d}^{(f)})\right] \rightarrow q^{(f)}$, with $q^{(f)}$ the standard charge of the fermion.  From  eq(\ref{eq:zanom}) and eq(\ref{NAcharge}) we conclude that anomalous transformations correspond to those with $\text{det}\left[ U(\bold{d}^{(f)}) \right] \neq 1$.

The mixed gravitational anomaly constraints are similarly straightforward and are given by:
\begin{align}
\label{eq:Dgravanom}
D-g-g&: \,\,\,\,\,\,\,\,\,\, \underset{\bold{d}^{(f)}}{\sum} tr\left[\tau(\bold{d}^{(f)})\right] \overset{!}{=} 0 \,\, \text{mod}\,\, \frac{N}{2} \\
\label{eq:Zgravanom}
Z-g-g&: \,\,\,\,\,\,\,\,\,\, \underset{f}{\sum} q^{(f)} = \underset{m}{\sum} q^{(m)}\cdot \text{dim}\, \bold{R}^{(m)} \overset{!}{=} 0 \,\, \text{mod}\,\, \frac{N}{2} 
\end{align} 
where $\bold{R}^{(m)}$ denotes the representations of all internal symmetries and the sum is such that each representation $\bold{R}^{(m)}$ only appears once.

\subsection{Anomalies in the UTZ model}
Turning to our universal texture zero model, we observe from Table \ref{tab:Zcharges} that we only ever assign fields to the (anti-)triplet or trivial singlet representations.  Yet from Table \ref{tab:groupreps} we see that determinants over these representations are unit in $\Delta(27)$.  As the summation in eq(\ref{eq:zanom}) and eq(\ref{eq:Dgravanom}) is only over fields that are non-trivial with respect to both $D$ and $G$ (or just $D$ for the gravitational anomalies), and since the coefficients are always $\propto \text{det}(h)$, we can make a strong claim:  \emph{we are free of all anomalies from the triangles $D-G-G$ and $D-g-g$, regardless of the form of the gauge group $G$.}

This means that we only have to be concerned with $Z-G-G$ and $Z-g-g$ anomalies, yet these also turn out to be trivially met, given the effective theory we have outlined.  For one, in a non-supersymmetric model, the only contributing fermions are the triplets of quarks, charged leptons, and neutrinos.  These are not charged under the $Z_{N}$ shaping symmetry, and thus contribute a vanishing anomaly coefficient.  For the supersymmetric case we would in principle have to include the fermionic partners to the familons $\theta_{i}$, $S$, Higgs(es) $H_{u,d}$ and the additional $\Sigma$ multiplet. However, we expect these fields to be heavy at the relevant scale of our effective Lagrangians, and hence they already `contribute' to the massive state contributions on the RHS of our anomaly equations.\footnote{The Higgsinos are trivially charged under the family symmetries we employ and thus would not contribute regardless of the relevant scale.} In order to check for anomaly cancellation above the mass scale of these supersymmetric bosons one would also have to construct the full theory including Froggatt-Nielsen type messenger states, which is beyond the scope of our discussion.

\begin{figure}[t]
\centering
\hspace{2mm}
\includegraphics[scale=0.25]{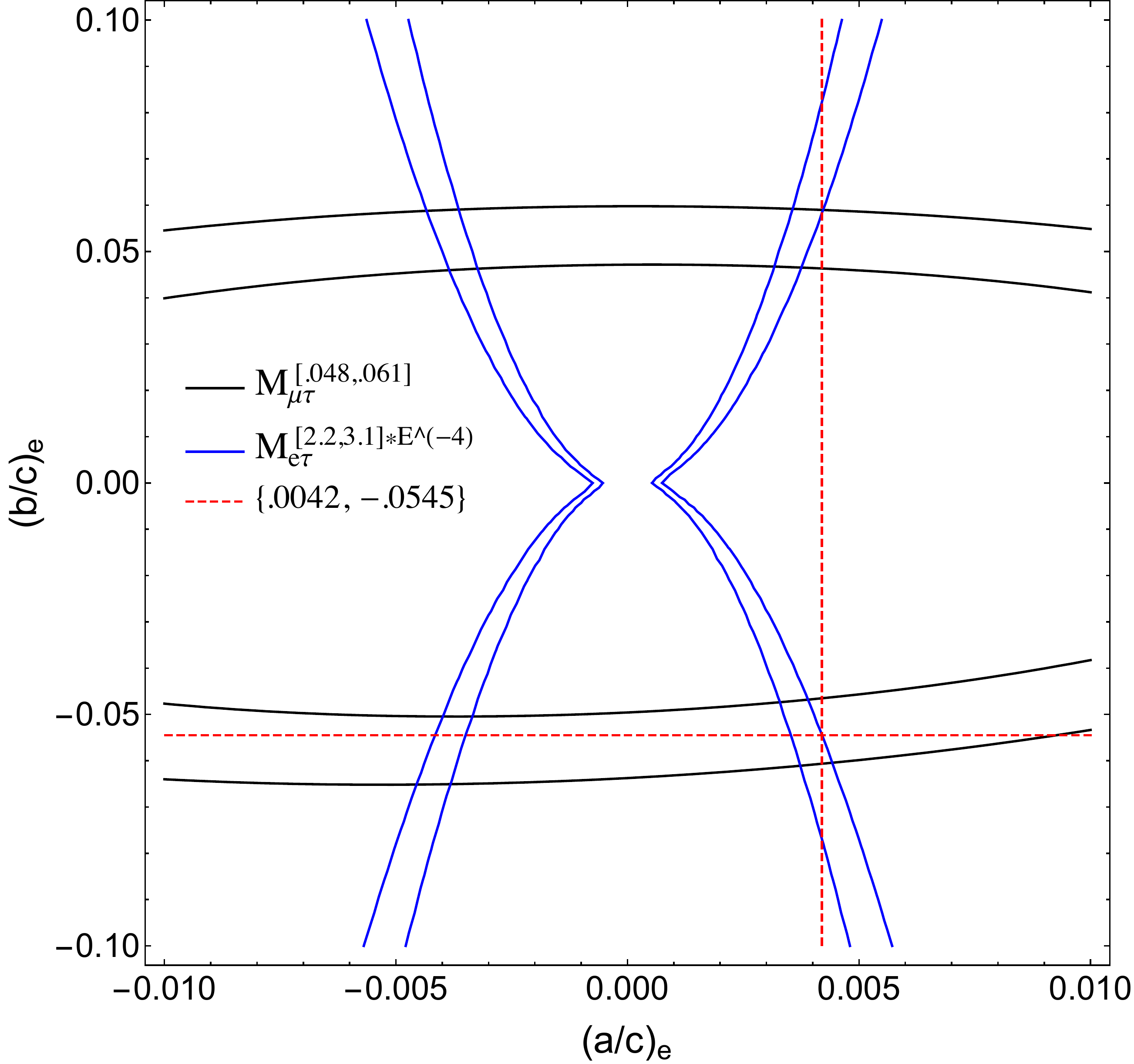}
\hspace{2mm}
\includegraphics[scale=0.26]{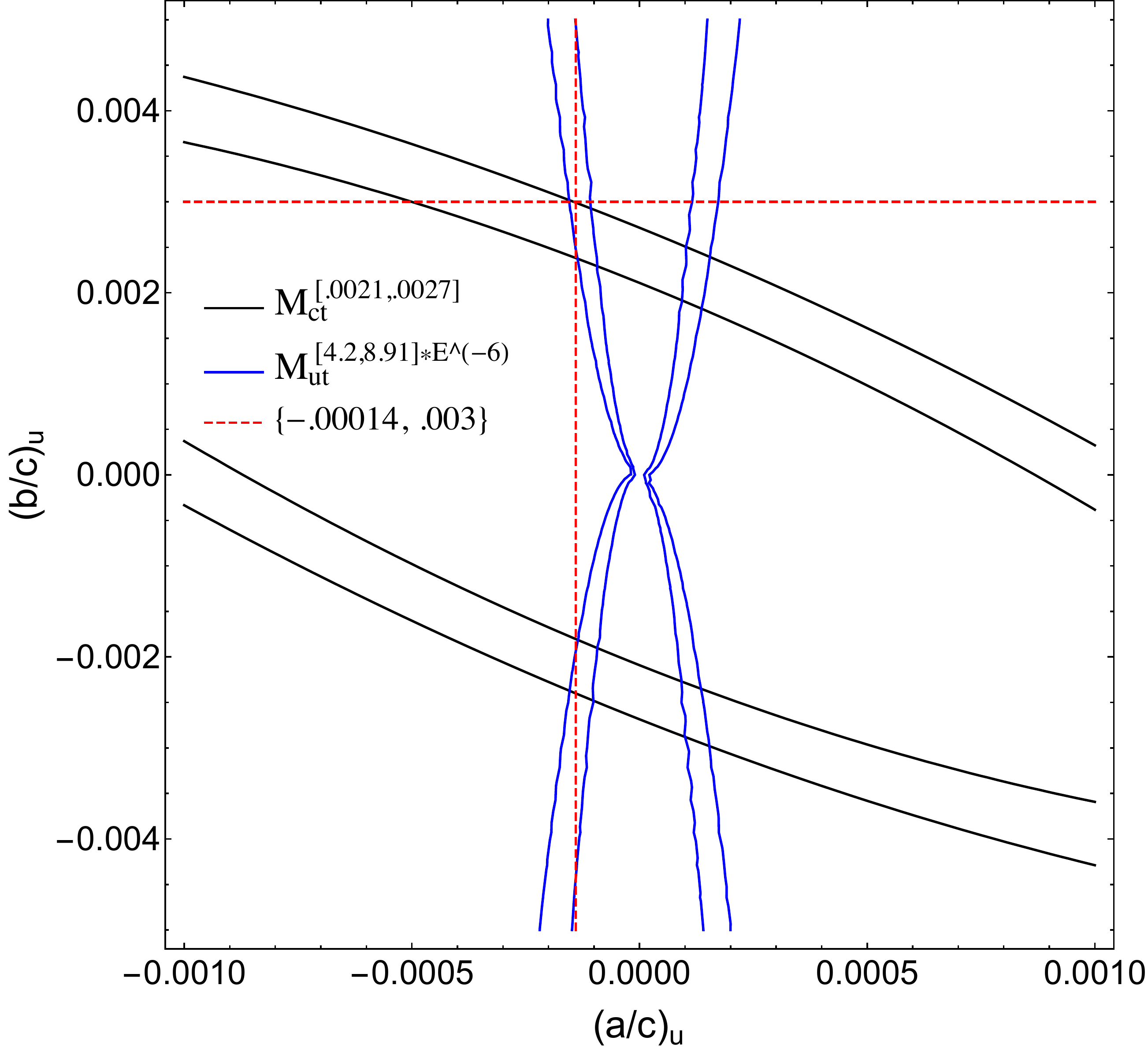} \\
\hspace{2mm}
\includegraphics[scale=0.27]{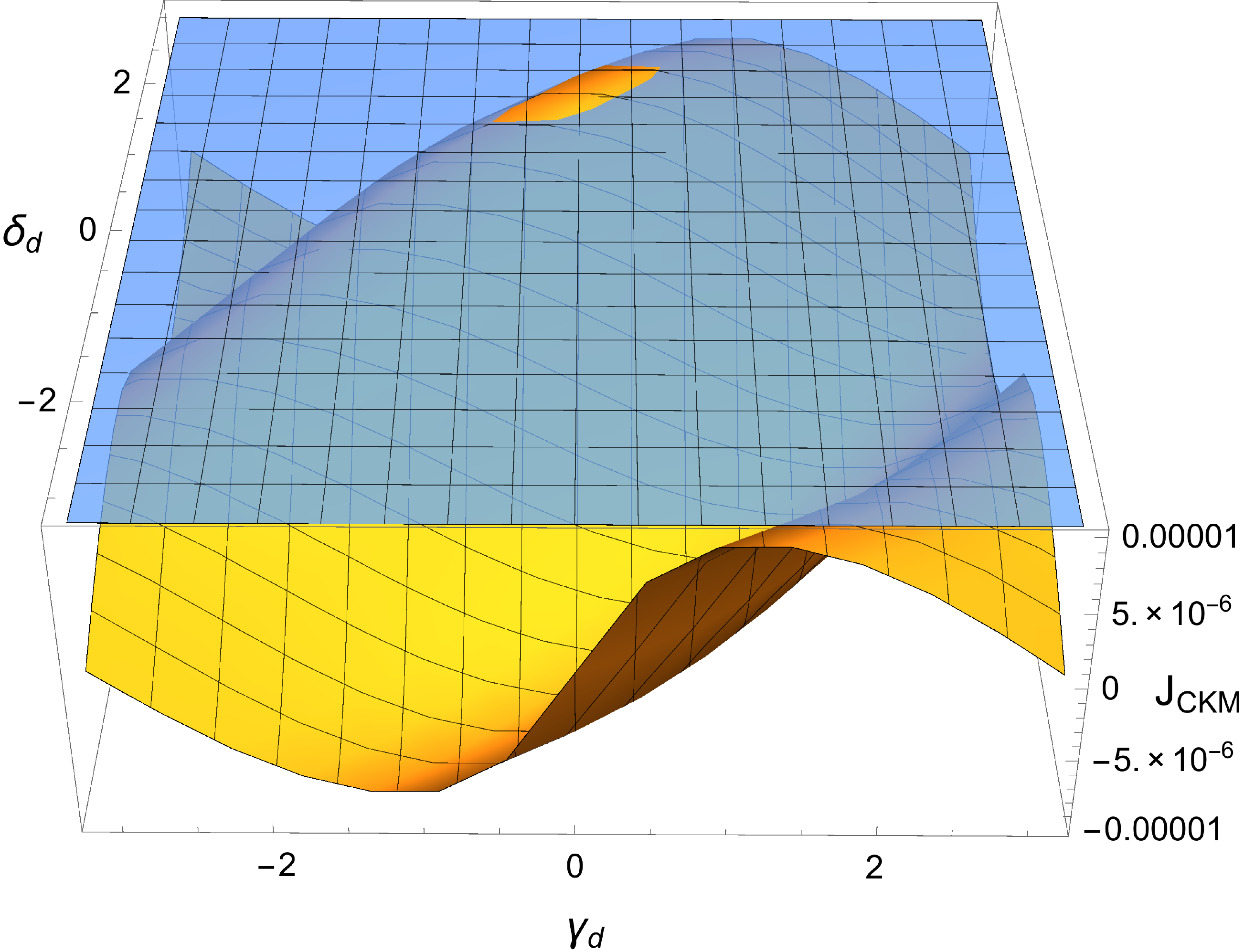}
\hspace{2mm}
\includegraphics[scale=0.235]{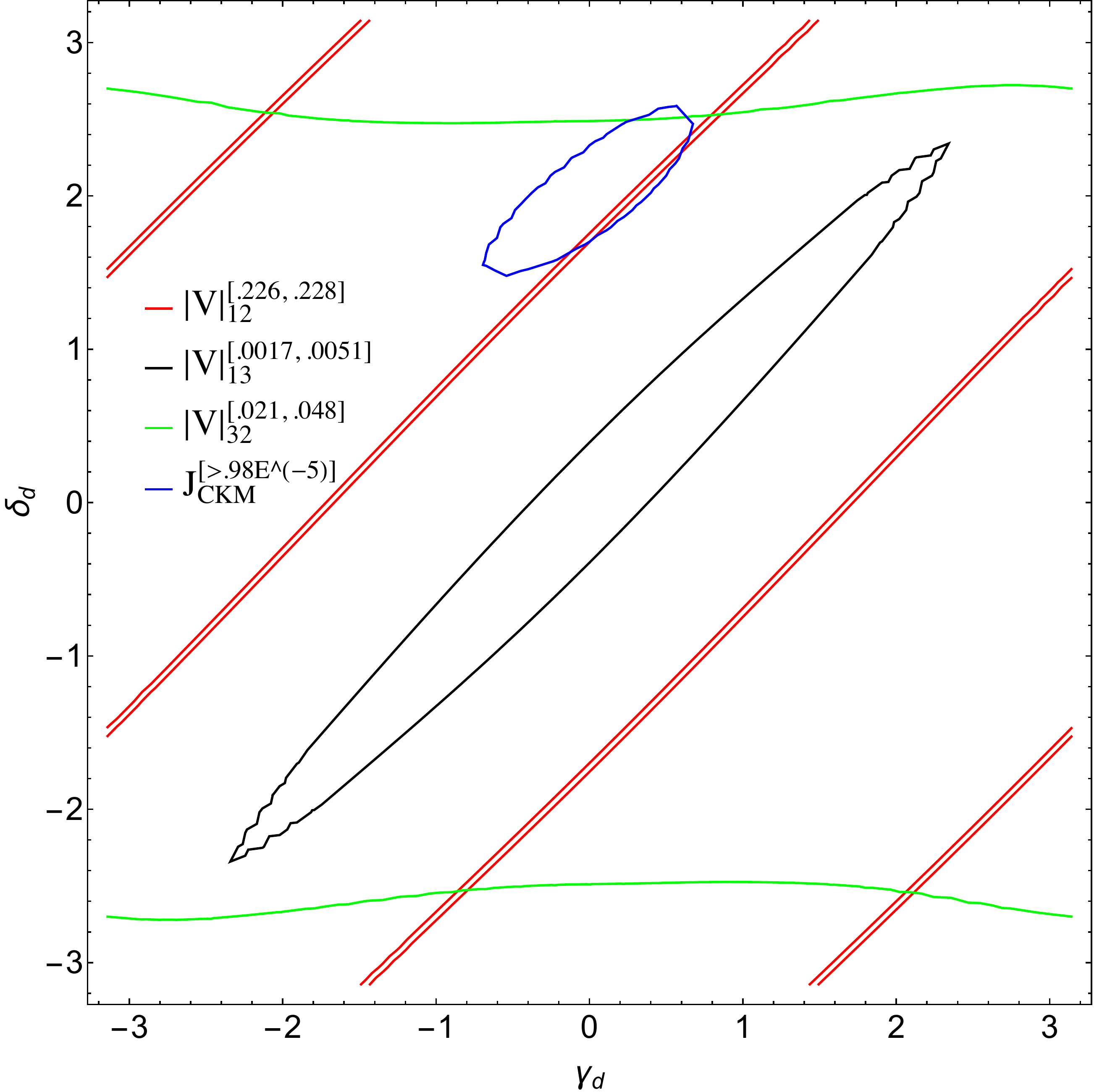}
\caption{Contours from our lowest order fit.  TOP LEFT:  Contours of the charged lepton mass fit.  Black contours represent the bounds for the ratio of $m_{\mu}/m_{\tau}$ whereas blue contours represent those for  $m_{e}/m_{\tau}$, both taken from \cite{Ross:2007az}.  The plot is at a fixed $m_{\tau}/m_{\tau} = 1$.  Red dashed lines represent our solution.  TOP RIGHT:  The same, but for up quarks.  BOTTOM LEFT:  The contours of the Jarlskog Invariant over the plane of the two free phases left in this fit after fixing the mass ratios $(a,b)_{e,u}$.  The blue plane represents the minimum $\mathcal{J}_{CKM}$ allowed in \cite{Ross:2007az}, and it is clear that portions of the parameter space (our solutions) can fit this.  BOTTOM RIGHT:  Contours of acceptable values of $|V_{ij}|_{CKM}$ and the CKM Jarlskog (interior of blue circle), also after fixing $(a,b)_{e,u}$.  The red line is the Cabibbo angle, and regions exterior to the black circle reflect acceptable values for the (1,3) element.  The relative magnitudes of the (1,3) and (3,1) elements are not successfully resolved at lowest order in our fit.  Higher order corrections as discussed in the text remedy this.}
\label{fig:FIT1cont}
\end{figure}

\section{Quantitative fit to the data}
\label{Pheno}
We now turn to a detailed numerical analysis of the associated phenomenology.  The core predictions of our model are complex symmetric mass matrices with a universal texture zero in the (1,1) position for all fermion families. 
As our model cannot determine the overall mass scale of the fermions, we work with matrices that have been rescaled by a factor from the (3,3) position that provides the bulk of the contribution to the third (heavy) generation.  For the Dirac masses, one obtains lowest order matrices of the form
\begin{equation}
\label{eq:MDIR}
\mathcal{M}_{i}^{D} \equiv \frac{M_{i}^{D}}{c} \simeq
\left(
\begin{array}{ccc}
0 & a\,e^{i(\alpha+\beta+\gamma)}  & a\,e^{i(\beta+\gamma)}  \\
a\,e^{i(\alpha+\beta+\gamma)}  & (b\,e^{-i\gamma}+2a\,e^{-i\delta})\,e^{i(2\alpha+\gamma + \delta)} & b\,e^{i(\alpha + \delta)}\\
a\,e^{i(\beta + \gamma)} & b\,e^{i(\alpha + \delta)} & 1 -  2 a\,e^{i\gamma} + b\,e^{i\delta}
\end{array}
\right)  
\end{equation}
where $i \in \lbrace u, d, e, \nu  \rbrace$ and where 
${a^{\prime}_i} = \frac{{{{\rm{v}}_{123}}{{\rm{v}}_{23}}\left\langle S \right\rangle }}{{\sqrt 6 M_{123,a}^3}},\quad {b^{\prime}_i} = \frac{{{r_a}{\rm{v}}_{23}^2\left\langle \Sigma  \right\rangle }}{{2M_{23,a}^3}},\quad {c_i} = \frac{{{\rm{v}}_3^2}}{{M_{3,a}^2}}$ and $r_{u,d,e,\nu}=(1,1,-3,-3)/3$.  The phases $\alpha$, $\beta$ are the those allowed from our generic complex vacuum alignment vectors while $\gamma$ and $\delta$ are the implicit phases of our complex mass matrix:
\begin{equation}
\label{eq:LOcont}
\frac{a^{\prime}}{c} = \vert \frac{a^{\prime}}{c} \vert \, e^{i \gamma} \equiv a \, e^{i \gamma}, \,\,\,\,\,\,\,\,\,\,\,\,\,\frac{b^{\prime}}{c} = \vert \frac{b^{\prime}}{c} \vert \, e^{i \delta} \equiv b \, e^{i \delta}
\end{equation}
The form of the mass matrix is the same for the heavy singlet Majorana neutrinos, but the overall mass scale is different.  We relabel the analogous free parameters as $a \rightarrow y$,  $b \rightarrow x$,  $c \rightarrow M$ (the mass scale in eq(\ref{MM})), $\gamma \rightarrow \rho$, $\delta \rightarrow \phi$, and keep the phases from vacuum alignment labeled as $\alpha$ and $\beta$.

\begin{table}
\centering
\renewcommand{\arraystretch}{1.35}
\begin{tabular}{|c||c|c|c|c||c|}
\hline
\multicolumn{6}{|c|}{Universal Texture Zero Input Parameters}\\
\hline
(1/c) $\times$ & $(a,b)_{e}$ & $(a,b)_{u}$  & $(a,b)_{\nu}$ & $(x,y)$ & $d_{d}$ \\
\hline
\text{L.O. Fit} & (.0042, -.0545)  & (-.00014, .003)& (4, 11.8)$\times 10^{-5}$ & (12.75, 4.055) $\times 10^{-13}$ & N.A.\\
\hline
\text{H.O. Fit} &  (.00416, -.0566) & (-.00014, .00275) & (4, 11.8)$\times 10^{-5}$ &  (12.75, 4.055) $\times 10^{-13}$ & .0145\\
\hline
\hline
& $(\gamma,\delta)_{e}$ & $(\gamma,\delta)_{u}$ & $(\gamma,\delta)_{\nu}$ & $(\rho,\phi)$ & $\psi_{d}$ \\
\hline
\text{L.O. Fit} & (.13, 1.83)  & (0, 0) & ($2\pi/5$,0) & (0,$-2\pi/5$) & N.A. \\
\hline
\text{H.O. Fit} & (0, 2) & (0,0) & ($2\pi/5$,0) & (0,$-2\pi/5$) & $\pi$\\
\hline
\end{tabular}
\caption{Free parameters used for fitting the fermionic mass and mixing spectrum.  As discussed in the text, only nine parameters are relevant to constraining the low-energy flavour phenomenology at lowest order in the operator product expansion.  The mass and phase parameters of the down quarks are implied by the corresponding values for the charged leptons. The parameters $d_{d}$ and $\psi_{d}$ are only relevant to fits including the higher order operator $\propto \Sigma\,S$ in eq(\ref{eq:HOLag}), which sources an independent entry analogous to eq(\ref{eq:LOcont}) labeled $d^{\prime}/c \equiv d e^{i \psi}$.  The subscript $d$ indicates that this contribution is only turned on for the down quarks (and hence also the charged leptons).  Note that the smallness of the Majorana neutrino parameters is compensated by a parameter determining their overall mass scale, which is not determined in our model.}
\label{tab:freeparameters}
\end{table}

In the quark and charged lepton sectors there are two mass ratios (a,b) and two phases ($\gamma$,$\delta$) for each family $(u,d,e)$, and an additional two phases ($\alpha$,$\beta$) from vacuum alignment.  This gives $(2+2)\times3 +2 = 14$ parameters, which reduces to 10 parameters if we assume an underlying GUT relation in the Georgi-Jarlskog form relating the down quarks to the charged leptons.  Six of these are phases, not all of which are physical.  In fact, only two phases are relevant at leading order \cite{Roberts:2001zy}, which we take to be $\gamma_{d}$ and $\delta_{d}$, leaving only six free parameters (including two phases).  Thus the 3 mixing angles and CP violating phase in the CKM matrix as well as the four quark and two charged lepton mass ratios are determined by just four real parameters and two phases.

The number of parameters needed in the neutrino sector is significantly reduced in the sequential limit where the $\nu_{3}^{c}$ exchange contribution to the see-saw masses is negligible.  There are just two parameters (including a phase) needed in this case (cf. eq(\ref{eq:2Dseesaw}-\ref{eq:heavyestate})), plus a parameter setting the scale of neutrino masses.  Thus, taking into account the contribution of the charged leptons, the leptonic mixing angles, atmospheric and solar mass differences, and CP violating phases are determined by two real parameters and a phase.  In summary, we see that both the charged fermion and neutrino sectors are over-constrained; 18 measurable quantities are determined by nine parameters, giving nine predictions at leading order in the operator expansion.

Having parameterized the mass matrices, one must then reliably calculate the associated mixing matrices.  The procedure we follow is enumerated below:
\begin{enumerate}
\item Find the matrix with columns as eigenvectors of $\mathcal{M}^{2} \equiv \mathcal{M} \cdot \mathcal{M}^{\dagger}$.
\item Diagonalize $\mathcal{M}$ by defining $\mathcal{\hat{M}} = U^{\dagger} \cdot \mathcal{M} \cdot U^{\star}$.
\item Define  $P = diag\left( e^{-i\,arg[\mathcal{\hat{M}}_{11}]/2},\,\,e^{-i\,arg[\mathcal{\hat{M}}_{22}]/2},\,\,e^{-i\,arg[\mathcal{\hat{M}}_{33}]/2} \right)$.  $U$ can now be made generic by $U \rightarrow U^{\prime} = U \cdot P$. 
\item Diagonalize the combination $\mathcal{M}^{2}$ by calculating  $U^{\prime \dagger} \cdot \mathcal{M}^{2} \cdot U^{\prime}$.
\item CKM matrices are now calculated as $V_{U}^{\dagger} \cdot V_{D}$, where $V = U^{\prime}$.
\item For the leptonic mixing the only thing that changes is that $\mathcal{M} \rightarrow \mathcal{M} \cdot \mathcal{M}^{M,-1}_{\nu_{R}} \cdot \mathcal{M}^{T}$ because of the see-saw.  Then  $V_{PMNS} = V_{e}^{\dagger} \cdot V_{\nu}$.  
\end{enumerate}
We note that this procedure is consistent with unitary rotations in the Standard Model Yukawa\footnote{Our low-energy neutrino mass term is of the Majorana form $\mathcal{L}_{\nu}^{M} \sim \bar{\nu}_{L} \, M_{\nu} \, \nu_{L}^{c}$.} sector and charged-current terms of the form:
\begin{align}
u^{I}_{L} &\rightarrow V_{U} u_{L} \,\,\,\,\,\,\,\,\,\, e^{I}_{L} \rightarrow V_{e} e_{L} \\
d^{I}_{L} &\rightarrow V_{D} d_{L} \,\,\,\,\,\,\,\,\,\, \nu^{I}_{L} \rightarrow V_{\nu} \nu_{L}
\end{align}
where $\lbrace u, d, e, \nu \rbrace_{L}$ are all left-handed family triplets.

\begin{table}[t]
\centering
\renewcommand{\arraystretch}{1.5}
\begin{tabular}{|c||c|c|c|c||c|c|c|c|}
\hline
\multicolumn{9}{|c|}{Uncertainties on UV Mixing Observables }\\
\hline
($\mu = M_{X}$)  & $\sin \theta_{12}^{q}$ & $\sin \theta_{23}^{q}$ &$\sin \theta_{13}^{q}$ & $\sin \delta_{CP}^{q}$ & $\sin \theta_{12}^{l}$ & $\sin \theta_{23}^{l}$ &$\sin \theta_{13}^{l}$ & $\sin \delta_{CP}^{l}$  \\
\hline
\text{Upper} & .228 & .0468  & .00508 & 1.000 & .588 & .800 & .155 & - \\ 
\hline
\text{Lower} & .226 & .0220 & .00169 & .186 & .520 & .620 & .139 & - \\
\hline
\hline
\multicolumn{9}{|c|}{Universal Texture Zero Mixing Predictions}\\
\hline
($\mu = M_{X}$)  & $\sin \theta_{12}^{q}$ & $\sin \theta_{23}^{q}$ &$\sin \theta_{13}^{q}$ & $\sin \delta_{CP}^{q}$ & $\sin \theta_{12}^{l}$ & $\sin \theta_{23}^{l}$ &$\sin \theta_{13}^{l}$ & $\sin \delta_{CP}^{l}$  \\
\hline
\text{L.O. Prediction} &  .226 & .0191 & .0042 & .561 & .554 & .778 & .152 & -.905\\
\hline 
\text{H.O. Prediction} & .226 & .0313 & .00307  & .788  & .543 & .751  & .153 & -.925 \\
\hline
\end{tabular}
\caption{TOP:  Uncertainty estimates for mixing observables in the UV.  For the quarks, the upper and lower values are estimated by taking overall error bands calculated by running the observables (with propagated experimental uncertainty) at various choices of $\tan \beta$ and other RGE input from \cite{Ross:2007az}.  We take the 3$\sigma$ global bounds from NuFit as extrema for the leptons, within which the leptonic CP violating phase is not constrained.  BOTTOM:  Predictions of mixing angles and CP violating phases extracted from the lower and higher order fits in eq\eqref{eq:VCKM1} - eq\eqref{eq:VPMNS2}.  Our predictions are within the estimated uncertainty bounds.}
\label{tab:mixingangles}
\end{table}

\subsection{Results of Numerical Fit}
\label{sec:FITS}
We have performed a fit where all of the up-quark phases are turned off and both $\gamma_{d}$ and $\delta_{d}$ are left free, as is consistent with \cite{Roberts:2001zy}.  This automatically also sets the corresponding phases for the charged leptons.  The values of all of the free parameters are given in Table \ref{tab:freeparameters}, and the corresponding predictions for the mass ratios, mixing angles, and CP violating phases are given in Tables \ref{tab:mixingangles}-\ref{tab:massfits} where we find excellent agreement with data (we use \cite{Ross:2007az} for our comparisons and do not assume specific values for $\tan \beta$, threshold corrections, etc.).  Contours of these predictions over planes representing our degrees of freedom are given in Figure \ref{fig:FIT1cont} for the charged leptons and up quarks assuming no higher order corrections as discussed in Section \ref{sec:HigherOrder}. The down quark contour is implied by the charged leptons.  Figure \ref{fig:FIT1cont} also includes contours for both the CKM Jarlskog and acceptable bands of CKM mixing after fixing $(a,b)_{e,u}$ but before fixing $\gamma_{d}$ and $\delta_{d}$.  

We also performed simple goodness-of-fit tests for both the `L.O' and `H.O Predictions' listed in Tables \ref{tab:mixingangles}-\ref{tab:massfits}, treating central values averaged between the listed uncertainty bands as our `observations.'  We find $\chi^{2}_{d.o.f.} < 1$ per degree of freedom in both cases \footnote{We do not include the leptonic CP violating phase in the fit as it is not constrained at the 3$\sigma$ level and thus constitutes a true prediction of the model.} despite the additional parameters involved in the H.O. fit, thus demonstrating the quality of our results.  We now discuss the quark and lepton sector mixings explicitly.

\subsubsection{CKM Matrix}
\label{sec:CKMfit}
The CKM mixing angles and Dirac CP-violating phase we predict, using only the lowest order parameters and applicable at the GUT scale, are given under `L.O. Predictions' in Table \ref{tab:mixingangles} and imply that the full CKM matrix and Jarlskog invariant are given by:
\begin{equation}
\label{eq:VCKM1}
|V_{CKM}|^{LO} =
\left(
\begin{array}{ccc}
.974 & .226  & .00420 \\
.226   & .974 & .0191\\
.00248  & .0194 & .9998
\end{array}
\right), \,\,\,\,\, \mathcal{J}_{CKM}^{LO} = 9.898\,\times 10^{-6}
\end{equation}
where it is clear that the Cabibbo sector in the (1,2) block is essentially perfect, the other off-diagonal elements are of the correct order of magnitude, and the Jarlskog invariant is successfully above its minimum value of $\sim9.8 \times 10^{-6}$.  On the other hand the (2,3) and (3,2) elements are a bit low and the (3,1) element is too small --- it should be approximately twice the (1,3) element.  While elements involving the third row or column are particularly sensitive to renormalization group running, being small, they are also sensitive to higher order corrections.  Thus it is of interest to determine whether these discrepancies can be eliminated by the leading higher order contribution discussed in Section \ref{sec:HigherOrder}.  We find that by only turning on the $d^{\prime}$ contribution to the down (and therefore also charged-lepton) mass matrices we can do so, but with the same number of free phases (we need one fewer from the lowest order parameter set).  Choosing $d_{d}  = .0145$ and its phase $\psi_{d} = \pi$, one achieves the mixing angles and CP phases listed under the `H.O. Prediction' of Table \ref{tab:mixingangles}, which implies the following CKM matrix and Jarlskog invariant: 
\begin{equation}
\label{eq:VCKM2}
|V_{CKM}|^{HO} =
\left(
\begin{array}{ccc}
.974 & .226  & .00307 \\
.226   & .974 & .0313\\
.00574  & .0309 & .9995
\end{array}
\right), \,\,\,\,\, \mathcal{J}_{CKM}^{HO} = 1.665\,\times 10^{-5}
\end{equation}

Note that the CKM elements, and in particular the relative magnitude of the (1,3) and (3,1) elements, are now in good agreement with data considering the uncertainties associated to the RG running to the GUT scale.  

\begin{table}[t]
\centering
\renewcommand{\arraystretch}{1.3}
\begin{tabular}{|c||c|c||c|c||c|c||c|}
\hline
\multicolumn{8}{|c|}{Uncertainties on UV Mass Ratios}\\
\hline
($\mu = M_{X}$) & $m_{e}/m_{\tau}$ & $m_{\mu}/m_{\tau}$ & $m_{u}/m_{t}$ &  $m_{c}/m_{t}$ & $m_{d}/m_{b}$ & $m_{s}/m_{b}$ & $\Delta m^{2}_{sol}/\Delta m^{2}_{atm}$\\
\hline 
\text{Upper} & .00031 & .061 & $8.91 \times 10^{-6}$ & .0027 & .0012 & .021 & .0336 \\
\hline
\text{Lower} & .00022 & .048 & $1.68 \times 10^{-6}$ & .00084 & .00035 & .008 & .021  \\
\hline
\hline
\multicolumn{8}{|c|}{Universal Texture Zero Mass Predictions}\\
\hline
($\mu = M_{X}$) & $m_{e}/m_{\tau}$ & $m_{\mu}/m_{\tau}$ & $m_{u}/m_{t}$ &  $m_{c}/m_{t}$ & $m_{d}/m_{b}$ & $m_{s}/m_{b}$ & $\Delta m^{2}_{sol}/\Delta m^{2}_{atm}$\\
\hline 
\text{L.O. Prediction} & .00031 & .055 & $7.16 \times 10^{-6}$ & .0027 & .00090  & .020 & .0213 \\
\hline 
\text{H.O. Prediction} & .00026 & .049 &  $7.89 \times 10^{-6}$ & .0025   & .0010 & .020 & .0213 \\
\hline
\end{tabular}
\caption{TOP:  Uncertainty estimates for mass ratios in the UV.  Bounds for the quarks are again taken from the running calculated in \cite{Ross:2007az} which includes a propagated experimental uncertainty, without assuming specific RGE input.  We estimate the neutrino mass squared difference in the UV from \cite{Antusch:2003kp}.  BOTTOM:   Predictions of mass ratios obtained from the numerical fits described in the text, both including (H.O.) and not including (L.O.) a higher order operator.  It is again clear that our predictions fit well within the uncertainty bounds.}
\label{tab:massfits}
\end{table}

\subsubsection{PMNS Matrix}
As discussed in Section \ref{neutrinos}, we expect the PMNS observables to be largely insensitive to RG running to the GUT scale, and so we wish to compare our results to the available NuFit data in eq\eqref{eq:Nufit}.  Taking only the lowest order parameter set, the leptonic mixing angles and Dirac CP phase are given under the `L.O. Prediction' of Table \ref{tab:mixingangles}, implying a PMNS matrix and Jarlskog invariant of
\begin{equation}
\label{eq:VPMNS1}
|V_{PMNS}|^{LO} =
\left(
\begin{array}{ccc}
.823 & .547  & .152 \\
.400   & .499 & .769 \\
.404 & .672 & .621
\end{array}
\right),\,\,\,\,\,\mathcal{J}_{PMNS}^{LO} = -.0304
\end{equation}
which is in excellent agreement with observation.  Of course, the PMNS sector is also sensitive to the higher order correction discussed above, which affects the charged lepton mixing matrix.  Upon turning on $d^{\prime}$, the predictions become:
\begin{equation}
\label{eq:VPMNS2}
|V_{PMNS}|^{HO} =
\left(
\begin{array}{ccc}
.830 & .536  & .153 \\
.405  & .534 & .742 \\
.384 & .654 & .652
\end{array}
\right),\,\,\,\,\,\mathcal{J}_{PMNS}^{HO} = -.0311
\end{equation}
which is still in total agreement with eq\eqref{eq:Nufit}.  We conclude that our UTZ model realizes very successful predictions across the spectrum of fermionic mass and mixing data.
\subsection{Error Determination}
\label{sec:errors}
We have seen that there are nine predictions involving the eighteen measurable quantities in the leading order fit and it is, of course, of interest to determine the errors in these predictions.  However, mainly due to the sizeable uncertainties associated with the continuation of the quark and charged lepton observables to the GUT scale that depend on unknown structure above the electroweak scale, we cannot determine the errors reliably in these sectors.  The continuation to high scales is more reliable in the neutrino sector and, despite the fact it is also sensitive to the charged lepton sector, we have attempted to get a rough estimate of the errors on our predictions for PMNS observables.

Of special interest is the constraint on the Dirac CP violating phase, given that it is not strongly constrained by data at the present.  This phase is particularly sensitive to the phases in the neutrino sector:  $\gamma_{\nu}$, $\delta_{\nu}$, $\rho$ and $\phi$.  To preserve a reasonable value for the neutrino mass ratios there is a strong correlation needed, namely $\gamma_{\nu} = -\phi$.  The remaining three phases which determine the Majorana and Dirac CP violating phases are also constrained by the fit to the observables, and the resulting limitation on their values simultaneously limits the variation of the CP phases.  To illustrate this we show in Figure \ref{fig:3sigCP} the variation of the Dirac CP violating phase as $\gamma_{\nu} = - \phi$ is varied over its 3$\sigma$ allowed range, keeping $\rho$ and $\delta_{\nu}$ fixed.  This corresponds to a variation of the Dirac phase in the range $\sin \delta^{l}_{CP} \in \left(-1.0, -0.82 \right)$.  

Of course, a serious evaluation of the errors should involve the error correlation with the other phases and parameters, but this is beyond the scope of this analysis.  Also shown in Figure \ref{fig:3sigCP} are the variations of the mixing angles as $\gamma_{\nu}$ varies.  In accordance with the form of eqs(\ref{eq:13sumrule}-\ref{eq:12sumrule}), only $\sin \theta_{23}^{l}$ varies appreciably. 
\begin{figure}[t]
\centering
\hspace{2mm}
\includegraphics[scale=0.4]{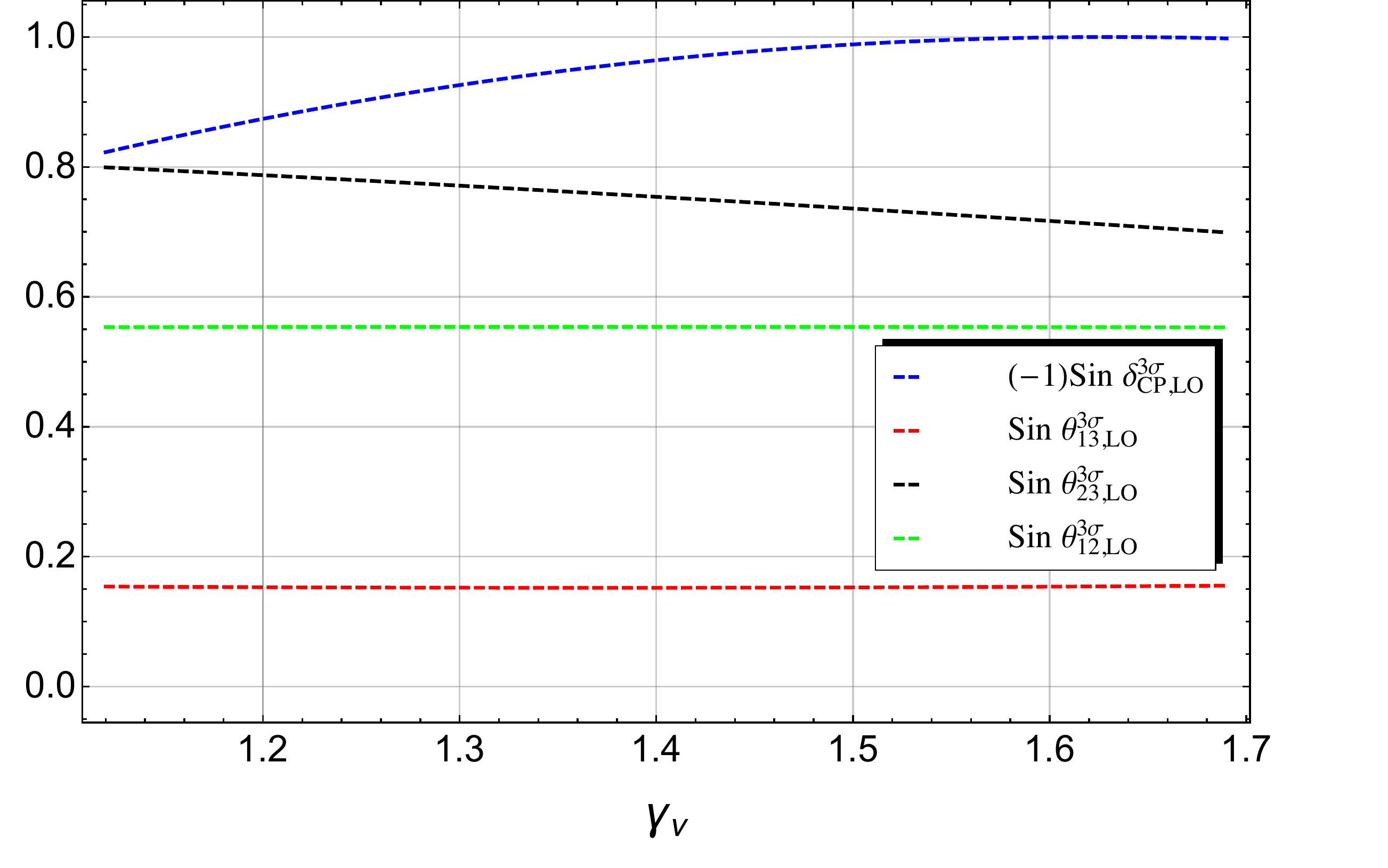}
\hspace{2mm}
\caption{Predictions obtained for $V_{PMNS}$ observables upon constraining the phase dependence of these observables with the $3\sigma$ data from NuFit and assumptions described in Section \ref{sec:errors}.  In this example we only utilize the L.O. UTZ Lagrangian.}
\label{fig:3sigCP}
\end{figure}

 \section{Summary and Conclusion}
 \label{sec:con}
 Most attempts to determine the pattern of fermion masses and mixings have assumed that there are separate symmetries describing the quark and the lepton sector in order to explain the disparate nature of quark and lepton mixing angles. However we have stressed that this may not be the case if the neutrino masses are generated by the see-saw mechanism. Exploiting this possibility we have constructed a viable model based on an egalitarian discrete symmetry model where all fermions and additional familons are triplets under the finite group, here $\Delta(27)$. As a result, the Dirac masses of both the quarks and leptons have the same form, albeit with different expansion parameters. The model is consistent with both an underlying stage of Grand Unification and the absence of discrete family symmetry anomalies.

A feature of the model is the appearance of a texture zero in the (1,1) position not only in the Dirac masses of all sectors, but also in both the heavy and light Majorana neutrino mass matrices. Combined with a symmetric mass matrix structure this leads to the successful Gatto-Sartori-Tonin relation for the Cabibbo angle. Assuming the Georgi- Jarlskog GUT structure for the down-quark and charged lepton mass matrices, the texture zero gives an excellent prediction for the electron mass. Finally in the neutrino sector the texture zero requires a departure from pure tribimaximal mixing, leading to a non-zero value for  $\theta^{l}_{13}$ consistent with the observed value.

In a detailed numerical analysis we show that the present measurements of fermion masses and mixings, up to the uncertainties in the radiative evolution of these parameters to the UV, are realized alongside of predictions for the Dirac leptonic CP violating phase. Overall, with just 9 free parameters, excellent agreement is found with the 18 observables in the charged fermion and neutrino sectors. As such it provides some evidence in favour of a dynamical rather than anarchical origin for fermion masses and mixings.

\section*{Acknowledgements}

IdMV acknowledges
funding from Funda\c{c}\~{a}o para a Ci\^{e}ncia e a Tecnologia (FCT) through the
contract IF/00816/2015.
This work was partially supported by Funda\c{c}\~ao para a Ci\^encia e a Tecnologia (FCT,
Portugal) through the project CFTP-FCT Unit 777 (UID/FIS/00777/2013) which is partially funded through POCTI (FEDER), COMPETE, QREN and EU. J.T. acknowledges research and travel support from DESY.  GGR thanks CERN for visiting support during which part of this work was conducted.

\appendix

\section{Vacuum alignment \label{VEV_align}}

\setcounter{equation}{0}  

 In what follows we consider the minimum number of triplet familon fields that can lead to the desired vacuum alignment. These are the four anti-triplet fields $\theta_{3,23,123}$ and $\theta$ introduced above together with a fifth triplet field $\theta_X$.  Assuming the underlying theory is supersymmetric we should include in the potential only those terms consistent with (spontaneously broken) SUSY.  For the case the associated familon superfields are R singlets there are no cubic terms in the superpotential involving only familon fields and hence, in the supersymmetric limit, no quartic terms. After SUSY breaking the scalar components of the superfields acquire SUSY breaking masses, giving the potential
 \be
 V_1(\theta_i)= m_i^2|\theta_i|^2
  \ee
  Radiative corrections can drive $m_i^2$ negative, triggering spontaneous breaking [45] of the family symmetry at a scale close to the scale at which $m_i^2$ is zero, and this may happen for all the familon fields.

These are the dominant terms that set the scale for the familon vevs. However, being $SU(3)_f$ invariant, these terms do not align the vevs in the manner required. To do that we need to consider terms allowed by the discrete symmetry that are not $SU(3)_f$ symmetric.  In studying this it is necessary to determine which couplings dominate.  In the context of a supersymmetric UV completion the leading quartic couplings come from F-terms associated with trilinear couplings to heavy mediators in the superpotential and, due to F-term decoupling, the couplings are small, suppressed by the square of the SUSY breaking scale over the mediator scale $(m_0/M)^2$, and depend sensitively on the mediator spectrum. As discussed above, we allow only triplet mediators and consider the most general set of effective couplings that can arise from the exchange of such mediators.

 Consider the case that  the dominant coupling for the $\theta_{3,123}$ fields is the self-coupling term
 \be
V_2(\theta_i)=h_i {\left( {{\theta _i}} \right)^2}{\left( {{\theta ^{\dag i}}} \right)^2}.
 \ee
 Minimising the potential\footnote{For clarity we assume real vevs here. The general case is presented below.} one sees that these terms align the field vevs, the direction depending on the sign of $h$:
\[\left\langle \theta_{i}  \right\rangle  = \left( {\begin{array}{*{20}{c}}
0\\
0\\
1
\end{array}} \right){{\rm{v}}_\theta },\;h_{i} < 0,\quad \left\langle \theta_{i}  \right\rangle  = \frac{1}{{\sqrt 3 }}\left( {\begin{array}{*{20}{c}}
1\\
1\\
{ 1}
\end{array}} \right){{\rm{v}}_\theta },\;h_{i} > 0\]
 These are in the directions required for $\theta_3$ and $\theta_{123}$!

To complete the model it is necessary to arrange the alignment of the $\theta _{23}$ field vev. The field $\theta_X$ can readily be made orthogonal to $\theta_{123}$ if its dominant effective coupling is
\be
V_3=k_1{\theta _{X,i}}\theta _{123}^{\dag i}{\theta _{123,j}}\theta_X ^{\dag j},\;\;k_1>0.
\ee
 However this term does not distinguish between $(0,1,-1)/\sqrt{2}$ and $(2,-1,-1)/\sqrt{6}$ (up to permutations of the elements). The latter vev is chosen if the dominant term
sensitive to the difference is
 \be
V_4=k_2 m_0 \theta_X^1\theta _X^2\theta _X^3
\ee
Although a cubic term in the superpotential involving the $\theta_X$  superfield is forbidden by R-symmetry, it is generated with coefficient $m_0$ after SUSY breaking. Then, in supergravity, the cubic term in the potential appears with $k_2=O(m_0/M)$ where $m_0$ is the gravitino mass.
 With this the final alignment of $\theta_{23}$ is driven by the term
\be
V_5=k_3\theta _{23,i}\theta _X^i\theta _{23}^{\dag j}\theta_X^{\dag j}+
k_4{\theta _{23,i}}\theta _{3}^{\dag i}{\theta _{3,i}}\theta_{23} ^{\dag i},\;\; \text{with}\,\, k_{3}>0\,\, \text{and}\,\, k_{4}<0
\ee
To summarise, the potential
\be
V = \sum\limits_{i = 3,123} {\left( {{V_1}({\theta _i}) + {V_2}({\theta _i})} \right)}  + {V_3} + {V_4} + {V_5}
\ee
aligns the fields in the directions
\be\left\langle {{\theta _3}} \right\rangle  = \left( {\begin{array}{*{20}{c}}
0\\
0\\
1
\end{array}} \right){{\rm{v}}_3},\quad \left\langle {{\theta _{123}}} \right\rangle  = \frac{1}{{\sqrt 3 }}\left( {\begin{array}{*{20}{c}}
e^{i\beta}\\
e^{i\alpha}\\
{ - 1}
\end{array}} \right){{\rm{v}}_{123}},\quad \left\langle {{\theta _{23}}} \right\rangle  = \frac{1}{{\sqrt 2 }}\left( {\begin{array}{*{20}{c}}
0\\
e^{i\alpha}\\
1
\end{array}} \right){{\rm{v}}_{23}},\quad \left\langle {{\theta ^{\dag}_X}} \right\rangle  = \frac{1}{{\sqrt 6 }}\left( {\begin{array}{*{20}{c}}
2e^{i\beta}\\
 - e^{i\alpha}\\
1
\end{array}} \right){{\rm{v}}_X}
\label{eq:finalvac}
\ee
where we have now included the relative phases explicitly. The vevs $\rm{v}_i$ may also be complex.
Note that further quartic terms allowed by the symmetries may be present but they should be subdominant to preserve this alignment. It is straightforwrd to assign a $Z_N$ charge to $\theta_X$ so that it does not contribute significantly to the fermion mass matrix.\footnote{A significant contribution of $\theta_{X}$ to fermion masses can also be avoided with an R-symmetry but, as this depends on the details of the underlying SUSY theory, we do not discuss this here.  Similarly, the cubic terms in the potential may determine some of the phases in eq(\ref{eq:finalvac}) but this too depends on the details of the symmetry properties of the underlying SUSY breaking sector.}

Finally,  it is necessary to align the $\theta$ familon that carries lepton number -1. This is readily the case through the potential
\be
{V_\theta } = {V_1}(\theta ) + {V_2}(\theta ) + k_5{\theta _{3,i}}\theta^{\dag i}{\theta _{i}}\theta _{3}^{\dag i},\;\;k_5<0
\ee
 
 \pagebreak


\begin{thebibliography}{9}
\bibitem{Lam:2007qc}
  C.~S.~Lam,
  Phys.\ Lett.\ B {\bf 656} (2007) 193
  doi:10.1016/j.physletb.2007.09.032
  [arXiv:0708.3665 [hep-ph]].


\bibitem{Altarelli:2005yp}
  G.~Altarelli and F.~Feruglio,
  Nucl.\ Phys.\ B {\bf 720} (2005) 64
  doi:10.1016/j.nuclphysb.2005.05.005
  [hep-ph/0504165].


\bibitem{Ma:2005qf}
  E.~Ma,
  Phys.\ Rev.\ D {\bf 73} (2006) 057304
  doi:10.1103/PhysRevD.73.057304
  [hep-ph/0511133].


\bibitem{Altarelli:2005yx}
  G.~Altarelli and F.~Feruglio,
  Nucl.\ Phys.\ B {\bf 741} (2006) 215
  doi:10.1016/j.nuclphysb.2006.02.015
  [hep-ph/0512103].


\bibitem{deMedeirosVarzielas:2005qg}
  I.~de Medeiros Varzielas, S.~F.~King and G.~G.~Ross,
  Phys.\ Lett.\ B {\bf 644} (2007) 153
  doi:10.1016/j.physletb.2006.11.015
  [hep-ph/0512313].


\bibitem{King:2017guk}
  S.~F.~King,
  Prog.\ Part.\ Nucl.\ Phys.\  {\bf 94} (2017) 217
  doi:10.1016/j.ppnp.2017.01.003
  [arXiv:1701.04413 [hep-ph]].


\bibitem{deAdelhartToorop:2011re}
  R.~de Adelhart Toorop, F.~Feruglio and C.~Hagedorn,
  Nucl.\ Phys.\ B {\bf 858} (2012) 437
  doi:10.1016/j.nuclphysb.2012.01.017
  [arXiv:1112.1340 [hep-ph]].


\bibitem{Lam:2012ga}
  C.~S.~Lam,
  Phys.\ Rev.\ D {\bf 87} (2013) no.1,  013001
  doi:10.1103/PhysRevD.87.013001
  [arXiv:1208.5527 [hep-ph]].


\bibitem{Holthausen:2012wt}
  M.~Holthausen, K.~S.~Lim and M.~Lindner,
  Phys.\ Lett.\ B {\bf 721} (2013) 61
  doi:10.1016/j.physletb.2013.02.047
  [arXiv:1212.2411 [hep-ph]].


\bibitem{Holthausen:2013vba}
  M.~Holthausen and K.~S.~Lim,
  Phys.\ Rev.\ D {\bf 88} (2013) 033018
  doi:10.1103/PhysRevD.88.033018
  [arXiv:1306.4356 [hep-ph]].


\bibitem{King:2013vna}
  S.~F.~King, T.~Neder and A.~J.~Stuart,
  Phys.\ Lett.\ B {\bf 726} (2013) 312
  doi:10.1016/j.physletb.2013.08.052
  [arXiv:1305.3200 [hep-ph]].


\bibitem{Lavoura:2014kwa}
  L.~Lavoura and P.~O.~Ludl,
  Phys.\ Lett.\ B {\bf 731} (2014) 331
  doi:10.1016/j.physletb.2014.03.001
  [arXiv:1401.5036 [hep-ph]].


\bibitem{Talbert:2014bda}
  J.~Talbert,
  JHEP {\bf 1412} (2014) 058
  doi:10.1007/JHEP12(2014)058
  [arXiv:1409.7310 [hep-ph]].


\bibitem{Joshipura:2014pqa}
  A.~S.~Joshipura and K.~M.~Patel,
  JHEP {\bf 1404} (2014) 009
  doi:10.1007/JHEP04(2014)009
  [arXiv:1401.6397 [hep-ph]].


\bibitem{Joshipura:2014qaa}
  A.~S.~Joshipura and K.~M.~Patel,
  Phys.\ Rev.\ D {\bf 90} (2014) no.3,  036005
  doi:10.1103/PhysRevD.90.036005
  [arXiv:1405.6106 [hep-ph]].


\bibitem{Yao:2015dwa}
  C.~Y.~Yao and G.~J.~Ding,
  Phys.\ Rev.\ D {\bf 92} (2015) no.9,  096010
  doi:10.1103/PhysRevD.92.096010
  [arXiv:1505.03798 [hep-ph]].


\bibitem{King:2016pgv}
  S.~F.~King and P.~O.~Ludl,
  JHEP {\bf 1606} (2016) 147
  doi:10.1007/JHEP06(2016)147
  [arXiv:1605.01683 [hep-ph]].


\bibitem{Varzielas:2016zuo}
  I.~de Medeiros Varzielas, R.~W.~Rasmussen and J.~Talbert,
  Int.\ J.\ Mod.\ Phys.\ A {\bf 32} (2017) no.06n07,  1750047
  doi:10.1142/S0217751X17500476
  [arXiv:1605.03581 [hep-ph]].


\bibitem{Yao:2016zev}
  C.~Y.~Yao and G.~J.~Ding,
  Phys.\ Rev.\ D {\bf 94} (2016) no.7,  073006
  doi:10.1103/PhysRevD.94.073006
  [arXiv:1606.05610 [hep-ph]].


\bibitem{deMedeirosVarzielas:2006fc}
  I.~de Medeiros Varzielas, S.~F.~King and G.~G.~Ross,
  Phys.\ Lett.\ B {\bf 648} (2007) 201
  doi:10.1016/j.physletb.2007.03.009
  [hep-ph/0607045].


\bibitem{Ma:2006ip}
  E.~Ma,
  Mod.\ Phys.\ Lett.\ A {\bf 21} (2006) 1917
  doi:10.1142/S0217732306021190
  [hep-ph/0607056].


\bibitem{Luhn:2007uq}
  C.~Luhn, S.~Nasri and P.~Ramond,
  J.\ Math.\ Phys.\  {\bf 48} (2007) 073501
  doi:10.1063/1.2734865
  [hep-th/0701188].


\bibitem{Varzielas:2015aua}
  I.~de Medeiros Varzielas,
  JHEP {\bf 1508} (2015) 157
  doi:10.1007/JHEP08(2015)157
  [arXiv:1507.00338 [hep-ph]].


\bibitem{Ishimori:2010au}
  H.~Ishimori, T.~Kobayashi, H.~Ohki, Y.~Shimizu, H.~Okada and M.~Tanimoto,
  Prog.\ Theor.\ Phys.\ Suppl.\  {\bf 183} (2010) 1
  doi:10.1143/PTPS.183.1
  [arXiv:1003.3552 [hep-th]].

\bibitem{Smirnov:1993af}
  A.~Y.~Smirnov,
  Phys.\ Rev.\ D {\bf 48} (1993) 3264
  doi:10.1103/PhysRevD.48.3264
  [hep-ph/9304205].

\bibitem{King:1998jw}
  S.~F.~King,
  Phys.\ Lett.\ B {\bf 439} (1998) 350
  doi:10.1016/S0370-2693(98)01055-7
  [hep-ph/9806440].


\bibitem{King:1999cm}
  S.~F.~King,
  Nucl.\ Phys.\ B {\bf 562} (1999) 57
  doi:10.1016/S0550-3213(99)00542-8
  [hep-ph/9904210].


\bibitem{King:1999mb}
  S.~F.~King,
  Nucl.\ Phys.\ B {\bf 576} (2000) 85
  doi:10.1016/S0550-3213(00)00109-7
  [hep-ph/9912492].


\bibitem{King:2002nf}
  S.~F.~King,
  JHEP {\bf 0209} (2002) 011
  doi:10.1088/1126-6708/2002/09/011
  [hep-ph/0204360].


\bibitem{Bjorkeroth:2015ora}
  F.~Björkeroth, F.~J.~de Anda, I.~de Medeiros Varzielas and S.~F.~King,
  JHEP {\bf 1506} (2015) 141
  doi:10.1007/JHEP06(2015)141
  [arXiv:1503.03306 [hep-ph]].


\bibitem{Bjorkeroth:2015uou}
  F.~Björkeroth, F.~J.~de Anda, I.~de Medeiros Varzielas and S.~F.~King,
  Phys.\ Rev.\ D {\bf 94} (2016) no.1,  016006
  doi:10.1103/PhysRevD.94.016006
  [arXiv:1512.00850 [hep-ph]].


\bibitem{Bjorkeroth:2017ybg}
  F.~Björkeroth, F.~J.~de Anda, S.~F.~King and E.~Perdomo,
  JHEP {\bf 1710} (2017) 148
  doi:10.1007/JHEP10(2017)148
  [arXiv:1705.01555 [hep-ph]].


\bibitem{Olechowski:1990bh}
  M.~Olechowski and S.~Pokorski,
  Phys.\ Lett.\ B {\bf 257} (1991) 388.
  doi:10.1016/0370-2693(91)91912-F


\bibitem{Ross:2007az}
  G.~Ross and M.~Serna,
  Phys.\ Lett.\ B {\bf 664} (2008) 97
  doi:10.1016/j.physletb.2008.05.014
  [arXiv:0704.1248 [hep-ph]].


\bibitem{Chiu:2016qra}
  S.~H.~Chiu and T.~K.~Kuo,
  Phys.\ Rev.\ D {\bf 93} (2016) no.9,  093006
  doi:10.1103/PhysRevD.93.093006
  [arXiv:1603.04568 [hep-ph]].


\bibitem{Gatto:1968ss}
  R.~Gatto, G.~Sartori and M.~Tonin,
  Phys.\ Lett.\  {\bf 28B} (1968) 128.
  doi:10.1016/0370-2693(68)90150-0


\bibitem{Georgi:1979df}
  H.~Georgi and C.~Jarlskog,
  Phys.\ Lett.\  {\bf 86B} (1979) 297.
  doi:10.1016/0370-2693(79)90842-6


\bibitem{Nilles:2012cy}
  H.~P.~Nilles, M.~Ratz and P.~K.~S.~Vaudrevange,
  Fortsch.\ Phys.\  {\bf 61} (2013) 493
  doi:10.1002/prop.201200120
  [arXiv:1204.2206 [hep-ph]].


\bibitem{deMedeirosVarzielas:2005ax}
  I.~de Medeiros Varzielas and G.~G.~Ross,
  Nucl.\ Phys.\ B {\bf 733} (2006) 31
  doi:10.1016/j.nuclphysb.2005.10.039
  [hep-ph/0507176].


\bibitem{Varzielas:2012ss}
  I.~de Medeiros Varzielas and G.~G.~Ross,
  JHEP {\bf 1212} (2012) 041
  doi:10.1007/JHEP12(2012)041
  [arXiv:1203.6636 [hep-ph]].


\bibitem{Nufit}
NuFIT 3.0 (2016), www.nu-fit.org

\bibitem{Esteban:2016qun}
  I.~Esteban, M.~C.~Gonzalez-Garcia, M.~Maltoni, I.~Martinez-Soler and T.~Schwetz,
  JHEP {\bf 1701} (2017) 087
  doi:10.1007/JHEP01(2017)087
  [arXiv:1611.01514 [hep-ph]].


\bibitem{Casas:1999tg}
  J.~A.~Casas, J.~R.~Espinosa, A.~Ibarra and I.~Navarro,
  Nucl.\ Phys.\ B {\bf 573} (2000) 652
  doi:10.1016/S0550-3213(99)00781-6
  [hep-ph/9910420].


\bibitem{Gupta:2014lwa}
  S.~Gupta, S.~K.~Kang and C.~S.~Kim,
  Nucl.\ Phys.\ B {\bf 893} (2015) 89
  doi:10.1016/j.nuclphysb.2015.01.026
  [arXiv:1406.7476 [hep-ph]].


\bibitem{Casas:1999ac}
  J.~A.~Casas, J.~R.~Espinosa, A.~Ibarra and I.~Navarro,
  Nucl.\ Phys.\ B {\bf 569} (2000) 82
  doi:10.1016/S0550-3213(99)00605-7
  [hep-ph/9905381].


\bibitem{Chankowski:1999xc}
  P.~H.~Chankowski, W.~Krolikowski and S.~Pokorski,
  Phys.\ Lett.\ B {\bf 473} (2000) 109
  doi:10.1016/S0370-2693(99)01465-3
  [hep-ph/9910231].


\bibitem{Antusch:2003kp}
  S.~Antusch, J.~Kersten, M.~Lindner and M.~Ratz,
  Nucl.\ Phys.\ B {\bf 674} (2003) 401
  doi:10.1016/j.nuclphysb.2003.09.050
  [hep-ph/0305273].


\bibitem{Hall:2013yha}
  L.~J.~Hall and G.~G.~Ross,
  JHEP {\bf 1311} (2013) 091
  doi:10.1007/JHEP11(2013)091
  [arXiv:1303.6962 [hep-ph]].


\bibitem{Krauss:1988zc}
  L.~M.~Krauss and F.~Wilczek,
  Phys.\ Rev.\ Lett.\  {\bf 62} (1989) 1221.
  doi:10.1103/PhysRevLett.62.1221


\bibitem{Ibanez:1991hv}
  L.~E.~Ibanez and G.~G.~Ross,
  Phys.\ Lett.\ B {\bf 260} (1991) 291.
  doi:10.1016/0370-2693(91)91614-2


\bibitem{Ibanez:1991wt}
  L.~E.~Ibanez and G.~G.~Ross,
  CERN-TH-6000-91.


\bibitem{Banks:1991xj}
  T.~Banks and M.~Dine,
  Phys.\ Rev.\ D {\bf 45} (1992) 1424
  doi:10.1103/PhysRevD.45.1424
  [hep-th/9109045].


\bibitem{Araki:2007zza}
  T.~Araki,
  Prog.\ Theor.\ Phys.\  {\bf 117} (2007) 1119
  doi:10.1143/PTP.117.1119
  [hep-ph/0612306].


\bibitem{Araki:2008ek}
  T.~Araki, T.~Kobayashi, J.~Kubo, S.~Ramos-Sanchez, M.~Ratz and P.~K.~S.~Vaudrevange,
  Nucl.\ Phys.\ B {\bf 805} (2008) 124
  doi:10.1016/j.nuclphysb.2008.07.005
  [arXiv:0805.0207 [hep-th]].


\bibitem{Csaki:1997aw}
  C.~Csaki and H.~Murayama,
  Nucl.\ Phys.\ B {\bf 515} (1998) 114
  doi:10.1016/S0550-3213(97)00839-0
  [hep-th/9710105].


\bibitem{Roberts:2001zy}
  R.~G.~Roberts, A.~Romanino, G.~G.~Ross and L.~Velasco-Sevilla,
  Nucl.\ Phys.\ B {\bf 615} (2001) 358
  doi:10.1016/S0550-3213(01)00408-4
  [hep-ph/0104088].
\end{thebibliography}
\end{document}